\documentclass[useAMS,usenatbib]{mn2e}
\usepackage{times,psfig,epsfig,graphics,graphicx,multicol,amssymb} 

\def\gca{GeCoA}
\def\aap{A\&A}
\def\apj{ApJ}

\def\apjl{ApJ}
\def\mnras{MNRAS}
\def\araa{ARA\&A}
\def\aj{AJ}

\def\physrep{Phys. Rep.}
\def\nat{Nat}

\def\lesssim{\mathrel{\hbox{\rlap{\hbox{\lower4pt\hbox{$\sim$}}}\hbox{$<$}}}}
\def\gesssim{\mathrel{\hbox{\rlap{\hbox{\lower4pt\hbox{$\sim$}}}\hbox{$>$}}}}

\begin{document}  
 \author[A. Morandi et al.]
{Andrea Morandi${}^1$\thanks{E-mail: andrea.morandi@studio.unibo.it}, Stefano Ettori${}^2$, 
Lauro Moscardini${}^{1,3}$\\
$^{1}$ Dipartimento di Astronomia, Universit\`a di Bologna, via Ranzani 1, 
I-40127 Bologna, Italy\\
$^{2}$ INAF-Osservatorio Astronomico di Bologna, via Ranzani 1, I-40127 Bologna, Italy \\
$^{3}$ INFN, Sezione di Bologna, viale Berti Pichat 6/2, I-40127 Bologna, Italy}

\date{}

\title[X-ray and SZ scaling relations]
{X-ray and Sunyaev-Zel'dovich scaling relations in galaxy clusters}

\pagerange{\pageref{firstpage}--\pageref{lastpage}} \pubyear{2007}
\maketitle

\label{firstpage}

\begin{abstract}
We present an analysis of the scaling relations between X-ray
properties and Sunyaev-Zel'dovich (SZ) parameters for a sample of 24
X-ray luminous galaxy clusters observed with Chandra and with measured
SZ effect. These objects are in the redshift range 0.14--0.82 and have
X-ray bolometric luminosity $L\ga 10^{45}$ erg
s$^{-1}$, with at least 4000 net counts collected for each source.  We
perform a spatially resolved spectral analysis and recover the density,
temperature $T$ and pressure profiles of the intra-cluster medium
(ICM), just relying on the spherical
symmetry of the cluster and the hydrostatic equilibrium hypothesis.
The combined analysis of the SZ and X-ray scaling relations is a
powerful tool to investigate the physical properties of the clusters
and their evolution in redshift, by tracing out their thermodynamical
history.  We observe that the correlations among X-ray quantities only
are in agreement with previous results obtained for samples of high-$z$ X-ray
luminous galaxy clusters.  On the relations involving SZ
quantities, we obtain that they correlate with the gas temperature
with a logarithmic slope significantly larger than the predicted value
from the self-similar model.  The measured scatter indicates, however,
that the central Compton parameter $y_0$ is a proxy of the gas
temperature at the same level of other X-ray quantities like
luminosity.  Our results on the X-ray and SZ scaling relations show a
tension between the quantities more related to the global energy of
the system (e.g. gas temperature, gravitating mass) and the indicators
of the structure of the ICM (e.g.  gas density profile, central
Compton parameter $y_0$).  Indeed, by using a robust fitting
technique, the most significant deviations from the values of the
slope predicted from the self-similar model are measured in the $L-T$,
$L-M_{\rm tot}$, $M_{\rm gas}-T$, $y_0-T$ relations.  When the slope
is fixed to the self-similar value, these relations consistently 
show a negative evolution suggesting a scenario in which the ICM at higher
redshift has lower both X-ray luminosity and pressure in the central
regions than the expectations from self-similar model.  These effects
are more evident in relaxed clusters in the redshift range 0.14-0.45,
where a more defined core is present and the assumed hypotheses on the
state of the ICM are more reliable.
\end{abstract} 

 
\section{Introduction}
 
Clusters of galaxies represent the largest virialized structures in
the present universe, formed at relatively late times.  The
hierarchical scenario provides a picture in which the primordial
density fluctuations generate proto-structures which are then subject
to gravitational collapse and mass accretion, producing larger and
larger systems.  The cosmic baryons fall into the gravitational
potential of the cluster dark matter (DM) halo formed in this way,
while the collapse and the subsequent shocks heat the intra-cluster
medium (ICM) up to the virial temperature ($0.5 \la T \la 10$ keV).

In the simplest scenario which neglects all non-radiative processes,
the gravity, which has not preferred scales, is the only responsible
for the physical properties of galaxy clusters: for this reason they
are expected to maintain similar properties when rescaled with respect
to their mass and formation epoch.  This allows to build a very simple
model to relate the physical parameters of clusters: the so-called
self-similar model \citep{1986MNRAS.222..323K,1991ApJ...383...95E}.
Based on that, we can derive scaling relations (see
Sect. \ref{scales1}) between X-ray quantities (like temperature $T$,
mass $M$, entropy and luminosity $L$), and between X-ray and
Sunyaev-Zel'dovich (SZ) measurements (like the Compton-$y$ parameter),
thanks to the assumption of spherical collapse for the DM halo and
hydrostatic equilibrium of the gas within the DM gravitational
potential.  These relations provide a powerful test for the adiabatic
scenario. In particular, in the recent years the studies about the
X-ray scaling laws \citep[see, e.g.,][]
{allen1998,1998ApJ...504...27M, 2004A&A...417...13E,
2005A&A...441..893A, 2005ApJ...628..655V, 2005ApJ...633..781K},
together with observations of the entropy distribution in galaxy
clusters \citep[see, e.g.,][]{1999Natur.397..135P,2003MNRAS.343..331P}
and the analysis of simulated systems including cooling and extra
non-gravitational energy injection \citep[see,
e.g.,][]{2004MNRAS.348.1078B} have suggested that the simple adiabatic
scenario is not giving an appropriate description of galaxy
clusters.  In particular the most significant deviations with respect to the
self-similar predictions are: (i) a lower (by $\sim 30-40$ per cent)
normalization of the $M-T$ relation in real clusters with respect to
adiabatic simulations \citep{1996ApJ...469..494E}; (ii) steeper slopes
for the $M-T$ and $L-T$ relations; (iii) an entropy ramp in the
central regions of clusters \citep[see, e.g.,][]{1999Natur.397..135P,
2003MNRAS.343..331P}.  These deviations are likely the evidence of
non-radiative processes, like non-gravitational heating due to energy
injection from supernovae, AGN, star formation or galactic winds
\citep[see, e.g.,][]{2001MNRAS.326..649P, 
2001ApJ...546...63T, 2001ApJ...555..597B,
2002MNRAS.330..329B, 2002MNRAS.336..409B, 2006ApJ...643..120B} or
cooling \citep[see, e.g.,][]{2000ApJ...544L...1B}.  More recently some
authors pointed out that there is a mild dependence of the X-ray
scaling relations on the redshift, suggesting that there should be an
evolution of these non-gravitational processes with $z$
\citep{2004A&A...417...13E}.

An additional and independent method to evaluate the role of radiative
processes is the study of the scaling relations based on the thermal
SZ effect \citep{1970CoASP...2...66S}, which offers a powerful tool
for investigating the same physical properties of the ICM, being the
electron component of cosmic baryons responsible of both the X-ray
emission and the SZ effect.  The advantage of the latter on the former
is the possibility of exploring clusters at higher redshift, because
of the absence of the cosmological dimming. Moreover, since the SZ
intensity depends linearly on the density, unlike the X-ray flux,
which depends on the squared density, with the SZ effect it is
possible to obtain estimates of the physical quantities of the sources
reducing the systematic errors originated by the presence of
sub-clumps and gas in multi-phase state and to study in a complementary
way to the X-ray analysis the effects of extra-physics on the collapse
of baryons in cluster dark matter halos, both via numerical
simulations \citep[]{2002ApJ...579...16W, 2004MNRAS.348.1401D, 2005MNRAS.356.1477D, 
2006ApJ...650..538N} and observationally (Cooray 1999; McCarthy et al. 2003a,b; Benson et al. 2004; LaRoque et al. 2006; Bonamente et al. 2006).

The main purpose of this paper is to understand how these SZ and X-ray
scaling relations evolve with redshift. In particular we want to
quantify how much they differ from the self-similar expectations in
order to evaluate the amplitude of the effects of the
non-gravitational processes on the physical properties of ICM.
Another issue we want to debate is which relations can be considered a
robust tool to link different cluster physical quantities: this has
important consequences on the possibility of using clusters as probes
for precision cosmology.  To do that, we have assembled a sample of 24
galaxies clusters, for which measurements of the Compton-$y$ parameter
are present in the literature. Respect the previous works we have done our 
own spatially resolved X-ray analysis recovering X-ray and SZ quantity 
necessary to investigate scaling relations. 
We have performed a combined spatial and spectral analysis
of the X-ray data, which allows us to derive the radial profile for
temperature, pressure, and density in a robust way.  These results, which have high spatial resolution,
rely only on the hydrostatic equilibrium hypothesis and spherical
geometry of the sources. Moreover we can compare the observed physical
quantities with the results of hydrodynamical numerical simulations in
a consistent way.

The paper is organized as follows.  In Sect. \ref{laoaw} we introduce
our cluster sample and we describe the method applied to determine the
X-ray properties (including the data reduction procedure) and the
corresponding SZ quantities.  In Sect. \ref{scales1} we report our
results about the scaling relations here considered, including the
presentation of the adopted fitting procedure.  Sect. \ref{cal} is
devoted to a general discussion of our results, while in
Sect. \ref{swqag} we summarize our main conclusions. We leave to the appendices the discussion of some tecnical details of our data reduction procedure. 

Hereafter we have assumed a flat $\Lambda CDM$ cosmology, with matter
density parameter $\Omega_{0m}=0.3$, cosmological constant density
parameter $\Omega_\Lambda=0.7$, and Hubble constant $H_{0}=70 \,{\rm
km/s/Mpc}$. Unless otherwise stated, we estimated the errors at the
68.3 per cent confidence level.

\section{The dataset}\label{laoaw}

\subsection{Data reduction}\label{laoa}

We have considered a sample of galaxy clusters for which we have SZ
data from the literature and X-ray data from archives (see Tables
{\ref{tab:2}} and {\ref{tab:1}}, respectively).  In particular, we
have considered the original sample of \cite{2003ApJ...591..526M}, to
which we added two more objects from the sample discussed by
\cite{2004ApJ...617..829B}. For all these clusters we have analyzed 
the X-ray data extracted from the {\it Chandra} archive. In total we
have 24 galaxy clusters with redshift ranging between 0.14 and 0.82,
emission-weighted temperature in the range 6-12 keV and X-ray
luminosity between $10^{45}$ and $10^{46}$ erg s$^{-1}$. 
In the whole sample we have 11 cooling core clusters and 13 no-cooling core 
ones (hereafter CC and NCC clusters, respectively) defined
according to the criterion that their cooling time in the inner 
regions is lower than the Hubble time at the cluster redshift.

\begin{table}
\begin{center}
\caption{
The SZ parameters for the galaxy clusters in our sample.  For each
object different columns report the name, the central value ($y_0$) of
the Compton $y$-parameter, the SZ flux integrated up to an
overdensity of $2500$ and over a fixed solid angle $\Omega=1$
arcmin ($y_{2500}$ and $y_\Omega$, respectively) divided by
the function $g(x)$ (see eq. \ref{eq:y}), and the parameter $\eta$ (see text). For two objects (namely
A1914 and RXJ2228+2037) the corresponding errors are not provided by
\protect \cite{2003ApJ...591..526M}:  in the following analysis  we will
assume for them a formal 1$\sigma$ error of 20 per cent.}
\begin{tabular}{l@{\hspace{.2em}} c@{\hspace{.2em}} c@{\hspace{.2em}} c@{\hspace{.2em}} c@{\hspace{.2em}} c@{\hspace{.2em}} c@{\hspace{.2em}} c@{\hspace{.2em}} c@{\hspace{.2em}} c@{\hspace{.2em}} c@{\hspace{.2em}} c@{\hspace{.2em}} c@{\hspace{.2em}} c@{\hspace{.2em}}}
\hline \
name  & $ y_{0}$ & $ y_{2500} $  &  $ y_{\Omega} $ & $\eta $\\
                &$(\times 10^{4})$&  (mJy)     &    (mJy)   &   \\
\hline \\
A1413 & ${1.61}_{-0.22}^{0.20}$ & ${40.3}\pm{5.2}$ & ${7.67}\pm{1.00}$ & 0.99 \\
A2204 & ${1.80}_{-0.62}^{0.46}$ & ${53.1}\pm{16.0}$ & ${7.91}\pm{2.38}$ & 0.79 \\
A1914 & ${1.59}_{..}^{..}$ & ${26.2}\pm{5.2}$ & ${6.41}\pm{1.28}$ & 1.20 \\
A2218 & ${1.37}_{-0.26}^{0.18}$ & ${25.7}\pm{4.1}$ & ${6.62}\pm{1.04}$ & 1.03 \\
A665 & ${1.37}_{-0.31}^{0.26}$ & ${37.1}\pm{7.7}$ & ${8.12}\pm{1.69}$ & 0.92 \\
A1689 & ${3.24}_{-0.20}^{0.22}$ & ${56.8}\pm{3.7}$ & ${13.34}\pm{0.86}$ & 0.94 \\
A520 & ${1.24}_{-0.19}^{0.17}$ & ${38.8}\pm{5.6}$ & ${7.53}\pm{1.08}$ & 1.10 \\
A2163 & ${3.56}_{-0.27}^{0.25}$ & ${142.6}\pm{10.5}$ & ${22.69}\pm{1.67}$ & 0.74 \\
A773 & ${2.37}_{-0.32}^{0.28}$ & ${34.8}\pm{4.4}$ & ${10.99}\pm{1.41}$ & 0.95 \\
A2261 & ${3.18}_{-0.40}^{0.35}$ & ${39.5}\pm{4.6}$ & ${12.40}\pm{1.46}$ & 0.92 \\
A2390 & ${3.57}_{-0.42}^{0.42}$ & ${78.4}\pm{9.2}$ & ${17.39}\pm{2.05}$ & 0.75 \\
A1835 & ${4.70}_{-0.29}^{0.31}$ & ${48.3}\pm{3.1}$ & ${16.44}\pm{1.06}$ & 0.80 \\
A697 & ${2.65}_{-0.32}^{0.32}$ & ${44.1}\pm{5.3}$ & ${13.88}\pm{1.66}$ & 0.96 \\
A611 & ${1.60}_{-0.24}^{0.24}$ & ${11.2}\pm{1.7}$ & ${5.39}\pm{0.82}$ & 1.02 \\
Zw3146 & ${1.61}_{-0.29}^{0.25}$ & ${15.8}\pm{2.6}$ & ${5.67}\pm{0.93}$ & 0.92 \\
A1995 & ${1.92}_{-0.16}^{0.14}$ & ${18.6}\pm{1.5}$ & ${8.22}\pm{0.65}$ & 1.06 \\
MS1358.4+6245 & ${1.47}_{-0.18}^{0.16}$ & ${13.4}\pm{1.5}$ & ${5.87}\pm{0.68}$ & 0.75 \\
A370 & ${2.36}_{-0.45}^{0.84}$ & ${18.9}\pm{5.2}$ & ${10.42}\pm{2.84}$ & 1.19 \\
RXJ2228+2037 & ${2.40}_{..}^{..}$ & ${14.9}\pm{3.0}$ & ${10.76}\pm{2.15}$ & 0.88 \\
RXJ1347.5-1145 & ${7.41}_{-0.68}^{0.63}$ & ${44.4}\pm{3.9}$ & ${19.60}\pm{1.74}$ & 0.70 \\
MS0015.9+1609 & ${2.33}_{-0.20}^{0.19}$ & ${11.5}\pm{1.0}$ & ${10.55}\pm{0.89}$ & 0.97 \\
MS0451.6-0305 & ${2.69}_{-0.19}^{0.17}$ & ${12.5}\pm{0.8}$ & ${9.04}\pm{0.60}$ & 1.31 \\
MS1137.5+6625 & ${1.53}_{-0.19}^{0.17}$ & ${2.4}\pm{0.3}$ & ${2.73}\pm{0.32}$ & 1.16 \\
EMSS1054.5-0321 & ${3.87}_{-1.12}^{1.19}$ & ${11.8}\pm{3.5}$ & ${13.93}\pm{4.16}$ & 1.04 \\
\hline \\ 
\end{tabular}

\label{tab:2}
\end{center}
\end{table}

We summarize here the most relevant aspects of the X-ray data
reduction procedure.  Most of the observations have been carried out
using ACIS--I, while for 4 clusters (A1835, A370, MS0451.6-0305,
MS1137.5+6625) we have data from the Back Illuminated S3 chip of
ACIS--S. We have reprocessed the event 1 file retrieved from the {\it
Chandra} archive with the CIAO software (version 3.2.2) distributed by
the {\it Chandra} X-ray Observatory Centre. 
We have run the tool {\tt aciss\_proces\_ events}
to apply corrections for charge transfer inefficiency (for the data at
153 K), re-computation of the events grade and flag background events
associated with collisions on the detector of cosmic rays.  We have
considered the gain file provided within CALDB (version 3.0) in this
tool for the data in FAINT and VFAINT modes. Then we have filtered the
data to include the standard events grades 0, 2, 3, 4 and 6 only, and
therefore we have filtered for the Good Time Intervals (GTIs)
supplied, which are contained in the {\tt flt1.fits} file. We checked
for unusual background rates through the {\tt script analyze\_ltcrv},
so we removed those points falling outside $\pm 3\sigma$ from the mean
value. Finally, we have applied a filter to the energy (300-9500 keV)
and CCDs, so as to obtain an events 2 file.

\subsection{Spatial and spectral analysis}\label{sp}
The images have been
extracted from the events 2 files in the energy range (0.5-5.0 keV), corrected by using the exposure map to remove the vignetting
effects, by masking out the point sources. So as to determine the centroid ($x_{\rm c},y_{\rm c}$) of the surface brightness we have fitted the images with a circular one-dimensional
(1D) isothermal $\beta$-model \citep[]{1976A&A....49..137C}, by adding a constant brightness model, and leaving $x_{\rm c}$ and $y_{\rm c}$ free as parameters in the best fit. We constructed a set of
$n$ ($n\sim 15-40$) circular annuli around the centroid of the surface brightness up to a maximum distance $R_{\rm
spat}$ (also reported in Table \ref{tab:1}), selecting the radii according
to the following criteria: the number of net counts of photons from
the source in the (0.5-5.0 keV) band is at least 200-1000 per annulus
and the signal-to-noise ratio is always larger than 2.  The background
counts have been estimated from regions of the same exposure which are
free from source emissions.
\begin{figure*}
\hbox{
\psfig{figure=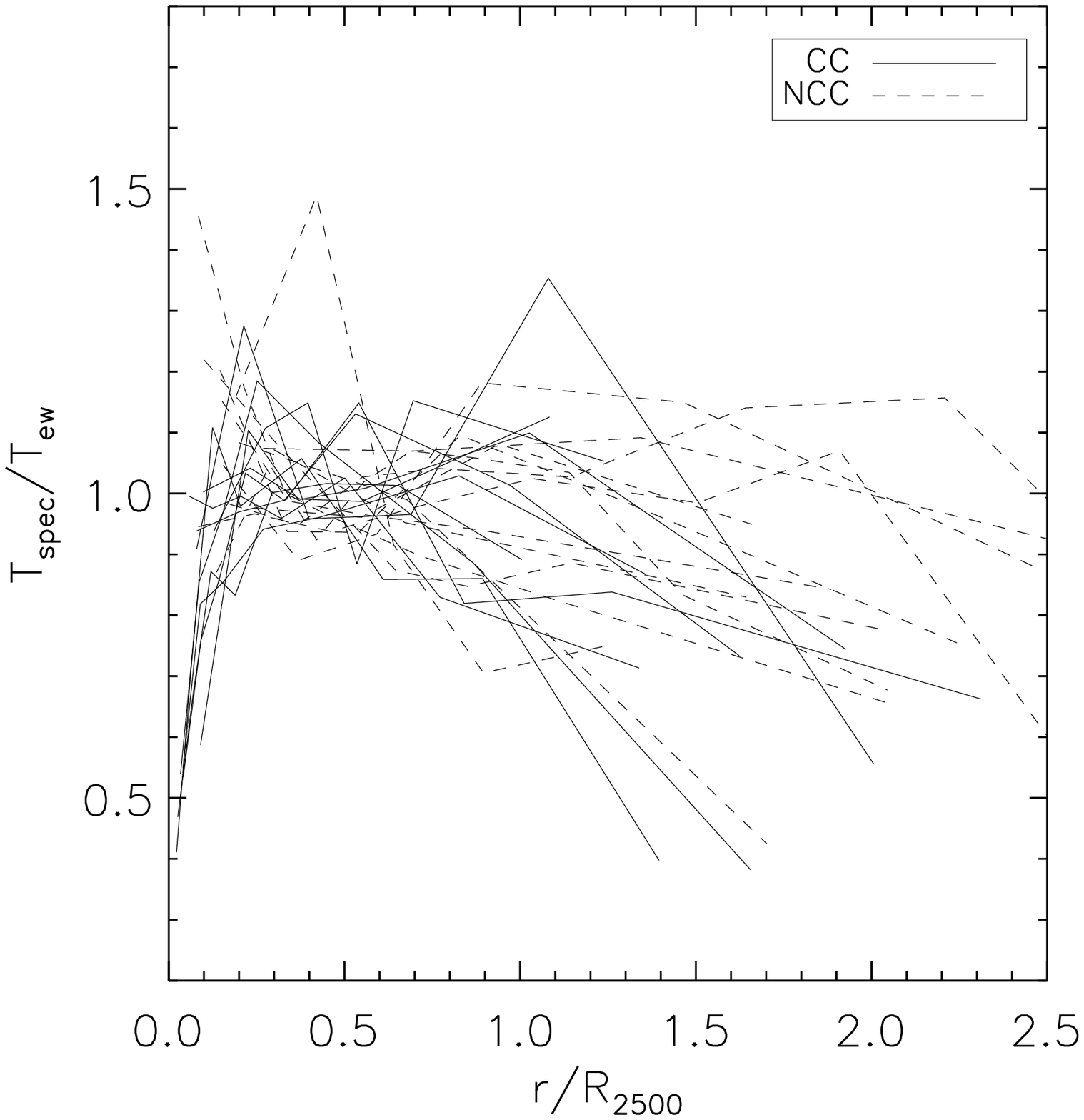 ,width=0.5\textwidth}
\psfig{figure=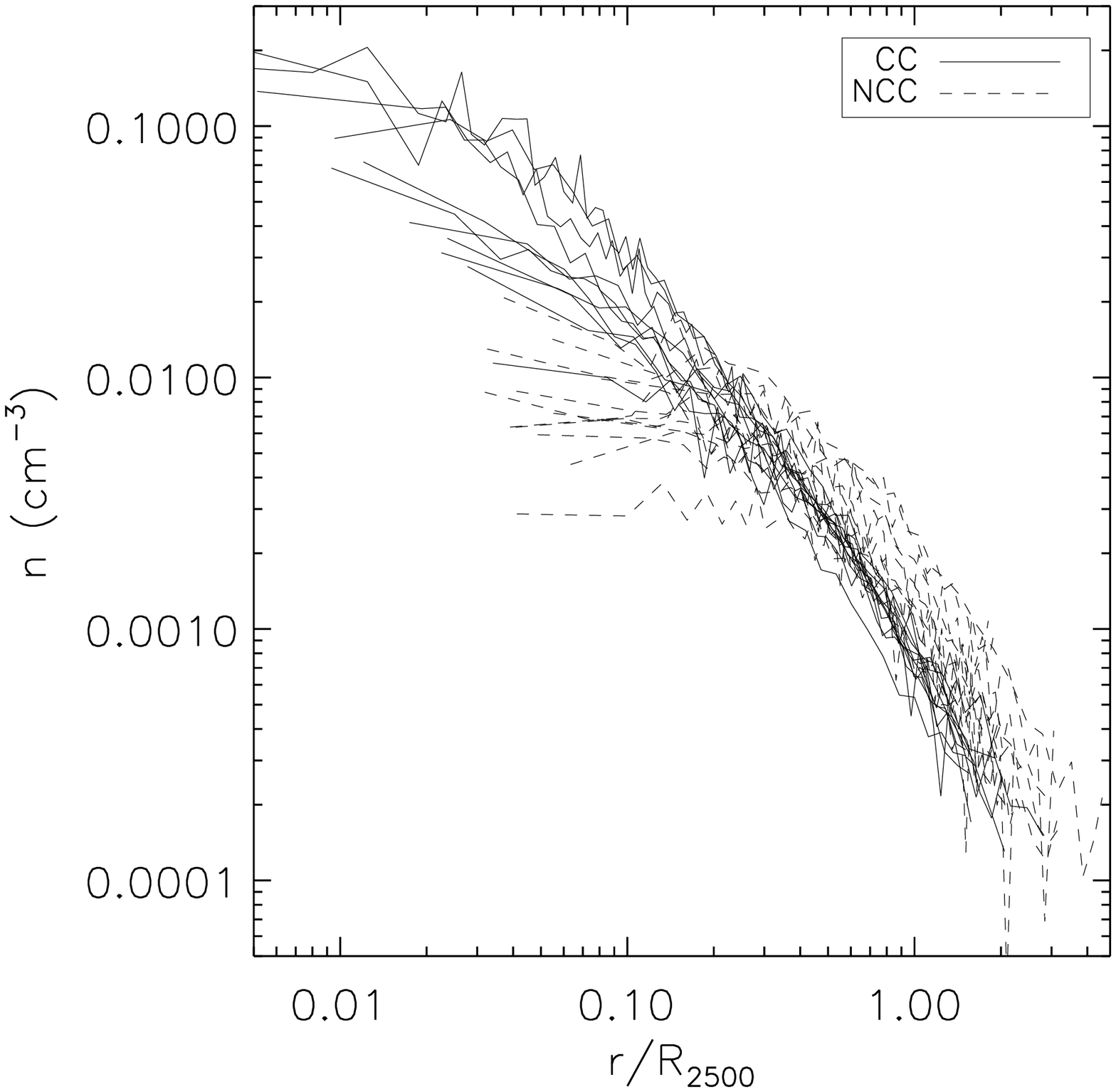 ,width=0.5\textwidth}
}
\caption[]{
The radial profiles for the projected temperature $T_{\rm spec}$,
normalized using the cooling-core corrected temperature $T_{\rm ew}$,
and for density are shown for all objects of our sample in the left
and right panels, respectively.  Solid and dashed lines refer to
clusters with or without a central cooling flow, respectively}
\label{fig:temp1} 
\end{figure*}

The spectral analysis has been performed by extracting the source
spectra from $n^*$ ($n^*\sim 3-8$) circular annuli of radius $r_{m}$ 
around the centroid of the surface
brightness.  We have selected the radius of each annulus out to a maximum
distance $R_{\rm spec}$ (reported in Table \ref{tab:1}), according to
the following criteria: the number of net counts of photons from the
source in the band used for the spectral analysis is at least 2000 per
annulus and corresponds to a fraction of the total counts always
larger than 30 per cent.
\begin{table*}
\begin{center}
\caption{
The X-ray properties of the galaxy clusters in our sample.  For each
object different columns report the name, the redshift $z$, the
identification number of the {\it Chandra} observation, the used ACIS
mode, the exposure time $t_{\rm exp}$, the neutral hydrogen absorption
$N_{\rm H}$ (the labels $f$ and $t$ refer to objects for which $N_{\rm
H}$ has been fixed to the Galactic value or thawed, respectively), the
physical scale corresponding to 1 arcmin, the maximum radii used for the spatial and for the spectral analysis ($R_{\rm spat}$ and $R_{\rm spec}$, respectively), the
emission-weighted temperature $T_{\rm ew}$, the metallicity $Z$ (in
solar units), a flag for the presence or not of a cooling core
(labeled CC and NCC, respectively), the mass-weighted temperature
$T_{\rm mw}$, the gas mass $M_{\rm gas}$, and the bolometric X-ray
luminosity $L$.  The last three columns refer to an overdensity of
$2500$. 
Sources extracted from the \protect \cite{2003ApJ...591..526M}
sample and from the \protect \cite{2004ApJ...617..829B} sample are
indicated by apices (1) and (2), respectively.}
\begin{tabular}{l@{\hspace{.3em}} c@{\hspace{.3em}} c@{\hspace{.3em}} c@{\hspace{.3em}} c@{\hspace{.3em}} c@{\hspace{.3em}} c@{\hspace{.3em}} c@{\hspace{.3em}} c@{\hspace{.3em}} c@{\hspace{.7em}} c@{\hspace{.3em}} c@{\hspace{.3em}} c@{\hspace{.7em}} c@{\hspace{.7em}} c@{\hspace{.7em}} c@{\hspace{.7em}} c@{\hspace{.7em}} c@{\hspace{.7em}} c@{\hspace{.7em}} c@{\hspace{.7em}} c@{\hspace{.7em}} c@{\hspace{.7em}} c@{\hspace{.7em}} c@{\hspace{.7em}}}
\hline \\
name   & $ z $ & obs. & ACIS & $ t_{\rm exp} $ & $ N_{\rm H} $ & $1'$ scale & $R_{\rm spat}$ & $R_{\rm spec}$  & 
$T_{\rm ew}$ & $Z$ & CC/ & $T_{\rm mw}$ & $M_{\rm gas}$ & $L$ \\
      & & &                 mode  & (ks)   &   $(10^{20} \mbox{cm}^{-2})$    &  (kpc)  &    (kpc)   &    (kpc)      
& (keV)       & ($Z_{\odot}$) & NCC & (keV) & ($10^{13}\,M_{\odot}$) & ($10^{45}$erg/s)\\
\hline \\
A1413$ ^{(1)}$ & 0.143 & 1661  & I & 9.7 & 2.2$({f})$ & 151 & 1111 & 1359 & $ 6.25_{-0.33}^{+0.36}$ & $0.45_{-0.10}^{+0.11}$ & CC & $  6.58 \pm   0.42$ & $   2.87 \pm 0.09$ & $1.28 \pm 0.03$ \\
A2204$ ^{(1)}$ & 0.152 & 6104  & I & 9.6 & 5.7$({f})$ & 159 & 1183 & 1262 & $ 9.18_{-0.65}^{+0.75}$ & $0.49_{-0.13}^{+0.14}$ & CC & $ 10.52 \pm   0.62$ & $   5.64 \pm 0.18$ & $4.21 \pm 0.14$ \\
A1914$ ^{(1)}$ & 0.171 & 3593  & I & 18.8 & 0.9$({f})$ & 175 & 1449 & 1576 & $ 8.93_{-0.45}^{+0.48}$ & $0.23_{-0.07}^{+0.07}$ & NCC & $  8.90 \pm   0.43$ & $   3.94 \pm 0.09$ & $1.88 \pm 0.05$ \\
A2218$ ^{(1)}$ & 0.176 & 1666  & I & 36.1 & 3.2$({f})$ & 179 & 1231 & 1320 & $ 6.88_{-0.30}^{+0.33}$ & $0.27_{-0.06}^{+0.06}$ & NCC & $  6.67 \pm   0.24$ & $   2.42 \pm 0.07$ & $0.84 \pm 0.02$ \\
A665$ ^{(1)}$ & 0.182 & 3586  & I & 29.1 & 4.2$({f})$ & 184 & 1589 & 1476 & $ 7.14_{-0.31}^{+0.33}$ & $0.28_{-0.06}^{+0.06}$ & NCC & $  7.02 \pm   0.20$ & $   2.61 \pm 0.08$ & $1.22 \pm 0.03$ \\
A1689$ ^{(1)}$ & 0.183 & 1663  & I & 10.6 & 1.8$({f})$ & 185 & 1446 & 1059 & $ 8.72_{-0.56}^{+0.63}$ & $0.23_{-0.10}^{+0.10}$ & CC & $  6.97 \pm   1.19$ & $   5.24 \pm 0.14$ & $3.15 \pm 0.09$ \\
A520$ ^{(1)}$ & 0.199 & 4215  & I & 66.2 & 3.5$({t})$ & 197 & 1327 & 1455 & $ 8.24_{-0.28}^{+0.31}$ & $0.32_{-0.05}^{+0.05}$ & NCC & $  9.70 \pm   0.55$ & $   3.47 \pm 0.09$ & $0.92 \pm 0.02$ \\
A2163$ ^{(1)}$ & 0.203 & 1653  & I & 71.1 & 17.5$({t})$ & 200 & 1846 & 1807 & $12.00_{-0.26}^{+0.28}$ & $0.24_{-0.03}^{+0.03}$ & NCC & $ 11.70 \pm   0.41$ & $   6.71 \pm 0.07$ & $4.80 \pm 0.05$ \\
A773$ ^{(1)}$ & 0.217 & 5006  & I & 19.8 & 1.4$({f})$ & 211 & 1105 & 1384 & $ 7.23_{-0.52}^{+0.62}$ & $0.37_{-0.12}^{+0.12}$ & NCC & $  7.09 \pm   0.36$ & $   2.34 \pm 0.11$ & $1.13 \pm 0.04$ \\
A2261$ ^{(1)}$ & 0.224 & 5007  & I & 24.3 & 3.3$({f})$ & 216 & 1588 & 1595 & $ 7.47_{-0.47}^{+0.53}$ & $0.37_{-0.10}^{+0.10}$ & CC & $  7.56 \pm   0.38$ & $   3.28 \pm 0.08$ & $2.02 \pm 0.07$ \\
A2390$ ^{(2)}$ & 0.232 & 4193  & S & 92.0 & 8.3$({t})$ & 222 & 1205 & 873 & $10.18_{-0.21}^{+0.23}$ & $0.29_{-0.03}^{+0.03}$ & CC & $ 10.02 \pm   0.16$ & $   6.98 \pm 0.08$ & $4.66 \pm 0.05$ \\
A1835$ ^{(1)}$ & 0.253 & 495  & S & 10.3 & 2.3$({f})$ & 237 & 914 & 970 & $ 8.62_{-0.54}^{+0.60}$ & $0.44_{-0.12}^{+0.12}$ & CC & $  8.75 \pm   0.80$ & $   5.89 \pm 0.60$ & $5.58 \pm 0.22$ \\
A697$ ^{(1)}$ & 0.282 & 4217  & I & 19.5 & 1.0$({t})$ & 256 & 1865 & 1679 & $10.21_{-0.75}^{+0.83}$ & $0.36_{-0.11}^{+0.11}$ & NCC & $  9.89 \pm   0.67$ & $   4.21 \pm 0.21$ & $2.52 \pm 0.09$ \\
A611$ ^{(1)}$ & 0.288 & 3194  & S & 35.1 & 5.0$({f})$ & 260 & 969 & 1172 & $ 6.06_{-0.34}^{+0.38}$ & $0.31_{-0.08}^{+0.09}$ & CC & $  6.32 \pm   0.37$ & $   2.46 \pm 0.07$ & $1.25 \pm 0.03$ \\
Zw3146$ ^{(1)}$ & 0.291 & 909  & I & 46.0 & 3.0$({f})$ & 262 & 1061 & 1287 & $ 7.35_{-0.26}^{+0.27}$ & $0.26_{-0.05}^{+0.05}$ & CC & $  8.48 \pm   0.30$ & $   5.56 \pm 0.15$ & $4.32 \pm 0.11$ \\
A1995$ ^{(1)}$ & 0.319 & 906  & S & 44.5 & 1.4$({f})$ & 279 & 877 & 914 & $ 7.56_{-0.41}^{+0.45}$ & $0.38_{-0.09}^{+0.09}$ & CC & $  7.75 \pm   0.48$ & $   3.39 \pm 0.11$ & $1.51 \pm 0.05$ \\
MS1358.4+6245$ ^{(1)}$ & 0.327 & 516  & S & 34.1 & 3.2$({t})$ & 283 & 796 & 813 & $ 7.51_{-0.61}^{+0.70}$ & $0.38_{-0.14}^{+0.15}$ & CC & $  8.05 \pm   0.58$ & $   2.98 \pm 0.15$ & $1.37 \pm 0.06$ \\
A370$ ^{(1)}$ & 0.375 & 515  & S & 48.6 & 3.1$({f})$ & 310 & 926 & 762 & $ 7.37_{-0.53}^{+0.58}$ & $0.28_{-0.10}^{+0.10}$ & NCC & $  7.73 \pm   0.41$ & $   3.35 \pm 0.14$ & $1.11 \pm 0.04$ \\
RXJ2228+2037$ ^{(1)}$ & 0.421 & 3285  & I & 19.8 & 4.9$({f})$ & 332 & 1320 & 1636 & $ 6.86_{-0.71}^{+0.89}$ & $0.35_{-0.15}^{+0.15}$ & NCC & $  7.48 \pm   0.81$ & $   2.36 \pm 0.15$ & $1.64 \pm 0.08$ \\
RXJ1347.5-1145$ ^{(1)}$ & 0.451 & 3592  & I & 57.7 & 4.9$({f})$ & 346 & 1558 & 1560 & $13.92_{-0.93}^{+1.14}$ & $0.19_{-0.09}^{+0.08}$ & CC & $ 15.32 \pm   0.83$ & $   8.99 \pm 0.19$ & $8.84 \pm 0.38$ \\
MS0015.9+1609$ ^{(1)}$ & 0.546 & 520  & I & 67.4 & 4.1$({f})$ & 383 & 1889 & 849 & $ 8.29_{-0.43}^{+0.49}$ & $0.32_{-0.06}^{+0.06}$ & NCC & $  8.00 \pm   0.37$ & $   3.13 \pm 0.09$ & $2.46 \pm 0.06$ \\
MS0451.6-0305$ ^{(1)}$ & 0.550 & 902  & S & 41.1 & 5.1$({f})$ & 385 & 1092 & 1325 & $ 9.09_{-0.61}^{+0.70}$ & $0.29_{-0.09}^{+0.10}$ & NCC & $  8.99 \pm   1.15$ & $   6.21 \pm 0.72$ & $3.92 \pm 0.12$ \\
MS1137.5+6625$ ^{(1)}$ & 0.784 & 536  & I & 116.4 & 3.5$({t})$ & 447 & 706 & 880 & $ 5.48_{-0.71}^{+0.89}$ & $0.25_{-0.22}^{+0.25}$ & NCC & $  6.28 \pm   0.57$ & $   1.73 \pm 0.10$ & $1.00 \pm 0.06$ \\
EMSS1054.5-0321$ ^{(2)}$ & 0.823 & 512  & S & 71.1 & 3.6$({f})$ & 455 & 763 & 895 & $ 9.00_{-1.10}^{+1.39}$ & $0.25_{-0.17}^{+0.17}$ & NCC & $  9.38 \pm   1.31$ & $   2.55 \pm 0.18$ & $1.35 \pm 0.13$ \\
\hline \\ 
\end{tabular}

\label{tab:1}
\end{center}
\end{table*}

The background spectra have been extracted from regions of the same
exposure in the case of the ACIS--I data, for which we always have
some areas free from source emission.  Conversely, for the ACIS--S
data we have considered the ACIS-S3 chip only and we have equally used
the local background, but we have checked for systematic errors due to
possible source contamination of the background regions. This is done
considering also the ACIS ``blank-sky" background files, which we have
re-processed if their gain file does not match the one of the events 2
file; then we have applied the aspect solution files of the
observation to the background dataset by using {\tt
reproject\_events}, so as to estimate the background for our data.  We
have verified that the spectra produced by the two methods are in good
agreement, and at last we decided to show only the results obtained
using the local background.

All the point sources has been masked out by visual inspection.  
Then we have calculated the redistribution
matrix files (RMF) and the ancillary response files (ARF) for each
annulus: in particular we have used the tools {\tt mkacisrmf} and {\tt
mkrmf} (for the data at 120 K and at 110 K, respectively) to calculate
the RMF, and the tool {\tt mkarf} to derive the ARF of the regions.

For each of the $n^*$ annuli the spectra have been 
analyzed by using the package XSPEC
\citep[]{1996ASPC..101...17A} after grouping the photons into
bins of 20 counts per energy channel (using the task {\tt grppha} from
the FTOOLS software package) and applying the $\chi^2$-statistics.
The spectra are fitted with a single-temperature absorbed MEKAL model
\citep{1992Kaastra, 1995ApJ...438L.115L} multiplied by a positive
absorption edge as described in \cite{2005ApJ...628..655V}: this
procedure takes into account a correction to the effective area
consisting in a 10 per cent decrement above 2.07 keV. The fit is
performed in the energy range 0.6-7 keV (0.6-5 keV for the outermost
annulus only) by fixing the redshift to the value obtained from
optical spectroscopy and the absorbing equivalent hydrogen column
density $N_{\rm H}$ to the value of the Galactic neutral hydrogen
absorption derived from radio data \citep{1990ARA&A..28..215D}, except
for A520, A697, A2163, MS1137.5+6625, MS1358.4+6245 and A2390, where
we have decided to leave $N_{\rm H}$ free due to the inconsistency
between the tabulated radio data and the spectral fit result.  Apart
for these objects where also the Galactic absorption is left free, we
consider three free parameters in the spectral analysis for $m-$th annulus: the
normalization of the thermal spectrum $K_{\rm m} 
\propto \int n^2_{\rm e}\, dV$, 
the emission-weighted temperature $T^*_{\rm proj,m}$; the metallicity 
$Z_{\rm m}$ retrieved by employing the solar abundance 
ratios from \cite{anders1989}.
The best-fit spectral parameters are listed in Table~\ref{tab:1}.

The total (cooling-core corrected) temperature $T_{\rm ew}$ has been
extracted in a circular region of radius $R$, with $100 {\rm \,
kpc}<R<R_{\rm spec}$, centred on the symmetrical centre of the
brightness distribution.  In the left panel of Fig. \ref{fig:temp1} we
present for all clusters of our sample the projected temperature
profile ($T_{\rm spec}$) normalized by $T_{\rm ew}$ as a function of
the distance from the centre $R$, given in units of $R_{2500}$, 
where $R_{2500}$ is the radius corresponding
to an ovedensity of $2500$.

\subsection{Spectral deprojection analysis}\label{depr}

To measure the pressure and gravitating mass profiles in our clusters,
we deproject the projected physical properties obtained with the
spectral analysis by using an updated and
extended version of the technique presented in 
\cite{2002A&A...391..841E} and discussed in full 
detail in Appendix~A. Here we
summarize briefly the main characteristics of the adopted technique:
(i) the electron density $n_e(r)$ is recovered both by deprojecting
the surface brightness profile and the spatially resolved spectral
analysis obtaining a few tens of radial measurements; 
(ii) once a functional form of the DM density profile
$\rho=\rho({\bf {r, q}})$, where ${\bf q}=(q_1, q_2, ... \, q_h)$ are
free parameters of the DM analytical model, and the gas pressure $P_0$ at $R_{spec}$ are assumed, the deprojected gas temperature, $T({\bf
q},P_0)$, is obtained by integration of the hydrostatic equilibrium
equation: 
\begin{equation}\label{aa2}
P({r},{\bf q},P_0)=P_0-\int_{R_{\rm spec}}^{r} n_{\rm gas}(s)\mu m_H \;{G
\;M({\bf q},s)\over{s^2}} \;d\,s\ ,
\end{equation}
where $\mu=0.6$ is the average molecular weight, $m_H$ is the proton mass. So $T({\bf q},P_0)=P({\bf q},P_0)/n_{\rm gas}$ expressed in keV units. 
In the present study, to parametrize the cluster mass distribution, we
consider two models: the universal density profile proposed by
\cite{1997ApJ...490..493N} (hereafter NFW) and the one suggested by
\cite{2004MNRAS.351..237R} (hereafter RTM).  

The NFW profile is given
by
\begin{equation}\label{nfwmodel}
\rho(x) = {{
\rho_{\rm c, z}\,\delta_{\rm c,NFW}} \over {  
({x/x_{\rm s}}) \left(1+{x/x_{\rm s}} \right)^2}}\ ,
\end{equation}
where $\rho_{\rm c, z}\equiv 3H(z)^2/ 8 \pi G$ is the critical density
of the universe at redshift $z$, $H_z\equiv E_z\,H_0$, $E_z \!=\left[\Omega_M (1+z)^3 +  (1-\Omega_M-\Omega_{\Lambda})(1+z)^2 + \Omega_{\Lambda}\right]^{1/2}$, and
\begin{equation}\label{aqrt}
\delta_{\rm c,NFW} = \frac{\Delta}{3} \frac{ c^3}{ {{\rm
ln}(1+c)-{c/(1+c)}}} \ ,
\end{equation}
where $c \equiv r_{\rm vir}/r_{\rm s}$ is the concentration parameter,
$r_{\rm s}$ is the scale radius, $x \equiv r/r_{\rm vir}$, $x_{\rm s}
\equiv r_{\rm s}/r_{\rm vir}$.

The RTM mass profile is given by:
\begin{equation}
\label{eq:bel}
\rho(x) = 
{{\rho_{\rm c, z}\, \delta_{\rm c,RTM}} \over{{x(x+{x_{s}^{*}})^{3/2}} }}\ ,
\end{equation}
with $x_{\rm s}^{*}\equiv r^{*}_{\rm s}/r_{\rm vir}$, where
$r^{*}_{\rm s}$ is a reference radius and $\delta_{\rm c,RTM}$ is
given by:
\begin{equation}
\label{eq:normbel}
\delta_{\rm c,RTM}\equiv 
\frac{\Delta}{6 \,\left[(1+2x_{\rm s}^{*})/ 
(1+x_{\rm s}^{*})^{1/2}-2 {x_{\rm s}^{*}}^{1/2} \right]}\ .
\end{equation}
So we have ${\bf q}=(c,r_{\rm s})$ and ${\bf q}=(x_{\rm s},r_{\rm 200})$ for the NFW and RTM models, respectively.
\begin{figure}
\psfig{figure=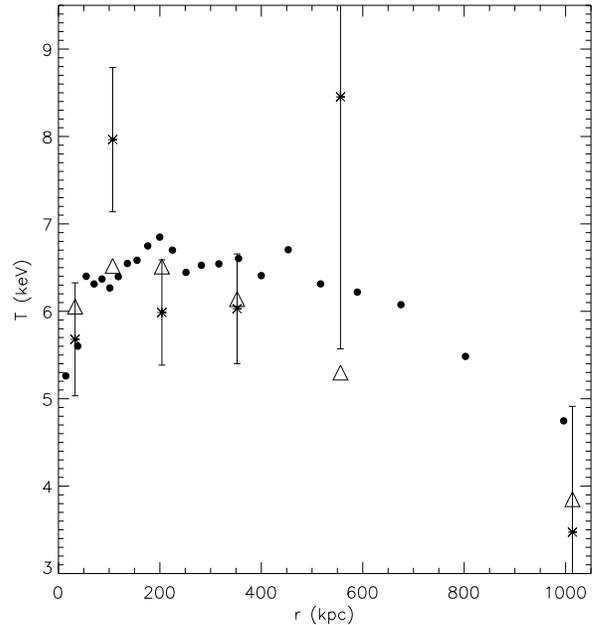,width=0.5\textwidth}
\caption[]{Example of temperature spectral deprojection for  cluster
A1413. We display the two quantities which enter in the eq. \ref{chi2ww} in the spectral deprojection analysis to retrieve the physical parameters: the observed spectral projected temperature 
$T^*_{\rm proj,m}$ (stars with errorbars) and 
the theoretical projected temperature (triangules, indicated as $T_{\rm proj,m}$ in Appendix~A). We also show the theoretical deprojected temperature $T({\bf q},P_0)$ (points), which generates $T_{\rm proj,m}$ through convenient projection tecniques.}
\label{fig:temp2}
\end{figure}

The comparison of the observed projected 
temperature profile $T^*_{\rm proj,m}$ 
(Sect. \ref{sp}) with the deprojected $T({\bf q},P_0)$ (eq. \ref{chi2ww} in Appendix~A), 
once the latter has been re-projected by correcting
for the temperature gradient along the line of sight as suggested in
\cite{2004MNRAS.354...10M}, provides the best estimate of the free
parameters $({\bf q},P_0)$ through a $\chi^2$ minimization, and therefore of 
$T({\bf q},P_0)$ (see an example in Figure \ref{fig:temp2}).

In the right panel of Fig. \ref{fig:temp1} we present the 
density profiles (plotted versus $r/R_{2500}$) as determined through
the previous method.  In general, we find there is 
no significant effect on the determination 
of the physical parameters when adopting the two different DM models.
Hereafter we will use the physical parameters
determined using the RTM model, reported with their
corresponding errors in Table \ref{tab:1}, where we also list the
exposure time, the number and the instrument (ACIS--I or ACIS--S) used
for each of the {\it Chandra} observations.

Finally we computed the total mass enclosed in a sphere of radius
$R_{\Delta}$ as $M({\bf q})(<R_{\Delta})=\int_{0}^{R_{\Delta}}
{\rho({r},{\bf q}) \, dV}$
where the radius $R_{\Delta}$ corresponds to a given overdensity
$\Delta$: we considered the cases where the overdensity is equal to
$2500$ and $500$.
The values for masses and radii, together with the parameters $({\bf
q},P_0)$ for the RTM model, are reported in Table \ref{tab:RTM}.  The
errors on the different quantities represent the 68.3 per cent
confidence level and are computed by looking to the regions in the
parameter space where the reduction of $\chi^2$ with respect to its
minimum value $\chi^2_{\rm min}$ is smaller than a given threshold,
fixed according to the number of degrees of freedom d.o.f. \citep[see,
e.g.,][]{1992nrca.book.....P}.  Notice that we included in the
eq.(\ref{aa2}) the statistical errors related to measurement errors of
$n_{\rm gas}(r)$.

\begin{table*} 
\begin{center}
\caption{
Different physical properties for the clusters in our sample.  For
each object the different columns report the name, the minimum value
for $\chi^2$ (with the corresponding number of degrees of freedom
d.o.f.), the virial radius $r_{\rm vir}$, the reference scale $x_{\rm
s}$, the value of the pressure $P_0$, the mass and the radius
corresponding to an overdensity of $2500$ 
($M_{2500}$ and $R_{2500}$,
respectively), the mass and the radius corresponding to an overdensity
of $500$ 
($M_{500}$ and $R_{500}$, respectively).  All quantities are
derived by assuming the RTM model.}
\begin{tabular}{l@{\hspace{.7em}} c@{\hspace{.7em}} c@{\hspace{.7em}} c@{\hspace{.7em}} c@{\hspace{.7em}} c@{\hspace{.7em}} c@{\hspace{.7em}} c@{\hspace{.7em}} c@{\hspace{.7em}} c@{\hspace{.7em}} c@{\hspace{.7em}} }
\hline \\
name & $\chi^2_{\rm min}$(d.o.f.) & $r_{\rm vir}$ &  $x_s$ & $P_0$ & $M_{2500}$ & $R_{2500}$ &  $M_{500}$ & $R_{500}$ \\
     &               &(kpc) &  & ($10^{-12}$ erg cm$^{-3}$)  & ($10^{14}M_{\odot}$) & (kpc)   &
($10^{14}M_{\odot}$) & (kpc) \\   
\hline \\
A1413 & 5.29(3)
 & $1853 \pm 255$ & $0.11 \pm 0.05$ & $1.87\pm 0.48$ & 2.30$(\pm 0.46)$ & 520$(\pm 57)$ & 5.58$(\pm 2.25)$ & 1195$(\pm 232)$ \\
A2204 & 4.24(5)
 & $2840 \pm 357$ & $0.16 \pm 0.03$ & $0.73\pm 0.71$ & 6.74$(\pm 1.40)$ & 742$(\pm 91)$ & 19.12$(\pm 6.68)$ & 1796$(\pm 320)$ \\
A1914 & 2.29(5)
 & $1809 \pm 202$ & $0.03 \pm 0.03$ & $1.98\pm 0.64$ & 3.39$(\pm 0.76)$ & 586$(\pm 61)$ & 5.99$(\pm 2.28)$ & 1212$(\pm 202)$ \\
A2218 & 0.61(2)
 & $1653 \pm 180$ & $0.06 \pm 0.04$ & $1.53\pm 0.33$ & 2.14$(\pm 0.30)$ & 502$(\pm 36)$ & 4.36$(\pm 1.33)$ & 1088$(\pm 152)$ \\
A665 & 1.23(5)
 & $2177 \pm 359$ & $0.31 \pm 0.16$ & $1.72\pm 0.44$ & 1.88$(\pm 0.16)$ & 480$(\pm 30)$ & 7.77$(\pm 2.71)$ & 1317$(\pm 260)$ \\
A1689 & 0.68(4)
 & $2159 \pm 323$ & $0.09 \pm 0.04$ & $1.24\pm 1.05$ & 4.14$(\pm 0.95)$ & 624$(\pm 76)$ & 9.40$(\pm 3.65)$ & 1402$(\pm 260)$ \\
A520 & 0.08(3)
 & $2487 \pm 1340$ & $0.23 \pm 0.74$ & $1.58\pm 0.57$ & 3.72$(\pm 0.93)$ & 599$(\pm 102)$ & 12.66$(\pm 8.50)$ & 1540$(\pm 546)$ \\
A2163 & 3.00(5)
 & $4884 \pm 637$ & $1.46 \pm 0.46$ & $1.44\pm 0.26$ & 4.07$(\pm 0.69)$ & 616$(\pm 102)$ & 52.58$(\pm 10.72)$ & 2472$(\pm 396)$ \\
A773 & 1.38(2)
 & $1672 \pm 486$ & $0.09 \pm 0.14$ & $2.90\pm 0.51$ & 2.01$(\pm 0.45)$ & 485$(\pm 60)$ & 4.54$(\pm 3.49)$ & 1087$(\pm 388)$ \\
A2261 & 2.82(3)
 & $1851 \pm 191$ & $0.10 \pm 0.03$ & $0.87\pm 0.46$ & 2.68$(\pm 0.44)$ & 532$(\pm 46)$ & 6.17$(\pm 1.79)$ & 1201$(\pm 168)$ \\
A2390 & 23.08(4)
 & $3557 \pm 497$ & $0.41 \pm 0.11$ & $2.66\pm 0.81$ & 6.59$(\pm 0.64)$ & 716$(\pm 52)$ & 33.22$(\pm 9.88)$ & 2099$(\pm 373)$ \\
A1835 & 0.80(1)
 & $2259 \pm 413$ & $0.14 \pm 0.05$ & $5.92\pm 1.70$ & 4.09$(\pm 1.34)$ & 606$(\pm 113)$ & 10.95$(\pm 6.39)$ & 1439$(\pm 414)$ \\
A697 & 1.17(4)
 & $2251 \pm 1491$ & $0.21 \pm 0.72$ & $2.69\pm 2.77$ & 3.23$(\pm 0.88)$ & 554$(\pm 94)$ & 10.46$(\pm 8.47)$ & 1402$(\pm 588)$ \\
A611 & 0.81(3)
 & $1719 \pm 242$ & $0.12 \pm 0.05$ & $2.14\pm 0.59$ & 2.11$(\pm 0.45)$ & 480$(\pm 56)$ & 5.18$(\pm 2.15)$ & 1107$(\pm 222)$ \\
Zw3146 & 4.36(3)
 & $2984 \pm 403$ & $0.31 \pm 0.07$ & $1.26\pm 0.86$ & 5.41$(\pm 0.81)$ & 656$(\pm 68)$ & 22.50$(\pm 7.58)$ & 1804$(\pm 344)$ \\
A1995 & 3.05(2)
 & $2585 \pm 2042$ & $0.32 \pm 0.67$ & $3.41\pm 2.84$ & 3.51$(\pm 1.32)$ & 562$(\pm 149)$ & 14.96$(\pm 13.96)$ & 1558$(\pm 804)$ \\
MS1358.4+6245 & 0.60(1)
 & $2748 \pm 2134$ & $0.38 \pm 0.66$ & $3.67\pm 3.11$ & 3.62$(\pm 1.64)$ & 566$(\pm 184)$ & 17.37$(\pm 16.70)$ & 1633$(\pm 885)$ \\
A370 & 4.21(1)
 & $2195 \pm 822$ & $0.23 \pm 0.33$ & $0.06\pm 1.44$ & 3.10$(\pm 0.74)$ & 528$(\pm 82)$ & 10.58$(\pm 7.70)$ & 1359$(\pm 518)$ \\
RXJ2228+2037 & 0.12(2)
 & $1648 \pm 1164$ & $0.19 \pm 0.73$ & $3.11\pm 1.72$ & 1.59$(\pm 0.46)$ & 415$(\pm 90)$ & 4.90$(\pm 4.35)$ & 1033$(\pm 464)$ \\
RXJ1347.5-1145 & 3.58(5)
 & $2703 \pm 187$ & $0.13 \pm 0.02$ & $0.01\pm 0.15$ & 9.49$(\pm 1.26)$ & 744$(\pm 56)$ & 24.00$(\pm 4.75)$ & 1734$(\pm 170)$ \\
MS0015.9+1609 & 0.96(4)
 & $2129 \pm 433$ & $0.48 \pm 0.39$ & $0.24\pm 0.55$ & 1.72$(\pm 0.25)$ & 406$(\pm 50)$ & 9.75$(\pm 2.82)$ & 1237$(\pm 224)$ \\
MS0451.6-0305 & 0.14(5)
 & $2118 \pm 1915$ & $0.21 \pm 0.73$ & $3.63\pm 3.24$ & 3.68$(\pm 1.43)$ & 522$(\pm 130)$ & 11.89$(\pm 12.85)$ & 1320$(\pm 725)$ \\
MS1137.5+6625 & 2.12(1)
 & $1468 \pm 284$ & $0.17 \pm 0.13$ & $0.13\pm 0.47$ & 1.91$(\pm 0.48)$ & 382$(\pm 58)$ & 5.47$(\pm 2.79)$ & 928$(\pm 238)$ \\
EMSS1054.5-0321 & 0.03(1)
 & $3060 \pm 1666$ & $1.39 \pm 0.75$ & $5.58\pm 7.61$ & 2.17$(\pm 1.26)$ & 393$(\pm 168)$ & 27.21$(\pm 21.81)$ & 1560$(\pm 930)$ \\
\hline 
\end{tabular}

\label{tab:RTM}
\end{center}
\end{table*}

\subsection{Determination of the X-ray properties}\label{phy}
The bolometric X-ray luminosity $L(<R_{\Delta})$ has been calculated by 
correcting the observed luminosity $L(100\, {\rm kpc}<r<R_{\rm spec})$  
determined from the spectral analysis performed by XSPEC excluding the central
cooling region of 100 kpc (the
results are reported in Table \ref{tab:1}):
\begin{equation}
L(<R_{\Delta})\!\! = \!\!
L(100\, {\rm kpc}<r<R_{\rm spec}) \ \frac{\int_{0}^{x_{\Delta}}
(1+x^2)^{-3 \beta} x^2 dx}{K_{\rm corr}}\ ,
\label{eq:lumx2}
\end{equation}
where $x=r/r_c$, $x_{\Delta}=R_{\Delta}/r_c$, $r_c$ and $\beta$ are the 
best-fit parameters of the $\beta$-model on 
the image brightness, $K_{\rm corr}$ is
the normalization of the thermal spectrum 
drawn with XSPEC, and corrected for the emission from the spherical
source up to 10 Mpc intercepted by the line of sight: $K_{\rm 
corr}=\int_{x_0}^{x_1}
(1+x^2)^{-3\beta} x^2 dx + \int_{x_1}^{x_2} (1+x^2)^{-3\beta} x^2
(1-\cos \theta)dx - \int_{x_0}^{x_2} (1+x^2)^{-3\beta} x^2 (1-\cos
\theta^*) dx$, with $\theta=\arcsin({x_1}/x)$, $\theta^*=\arcsin({x_0}/x)$, 
$x_0=100\, {\rm kpc}/r_c$, $x_1=R_{\rm spec}/r_c$ and $x_2=10\, {\rm 
Mpc}/r_c$.

The gas mass $M_{\rm gas}$ enclosed in a circular region having
overdensity $\Delta$ has been computed from the total gas 
density $n_{\rm gas,j}$,
that we directly obtained from the spectral deprojection, up to
$R_{\rm spec}$. We have checked that the exclusion of
the central cooling region does not significantly affect the resulting
values for $M_{\rm gas}$.

Finally we have estimated the total mass-weighted temperature: 
\begin{equation}\label{tmw}
T_{\rm
mw} \equiv {\left({\sum_{i=1}^p T_j({\bf q},P_0) \,
m_i}\right)/{\sum_{i=1}^p m_i}}
\end{equation}
which can be compared to the total
emission-weighted temperature $T_{\rm ew}$; $p$ represents the number
of annuli inside $R_{2500}$. Notice that our average deprojected
temperature profile implies the following relation between the
maximum, the deprojected and the mass-weighted temperatures: $T_{\rm
max}:T_{\rm ew}:T_{\rm mw}=1:0.67:0.69$ ($1:0.83:0.88$ for the CC-only
subsample).  The physical parameters obtained in this way are also
listed in Table \ref{tab:1} for all clusters of our sample.

\subsection{Determination of the Sunyaev-Zel'dovich properties}\label{thsz0}

The thermal SZ \citep{1970CoASP...2...66S} effect is a very small
distortion of the spectrum of the cosmic microwave background (CMB),
due to the inverse Compton scatter between cold CMB photons and hot
ICM electrons \citep[for recent reviews see,
e.g.,][]{1999PhR...310...97B,2002ARA&A..40..643C,2005bmri.conf...57R}.
This comptonization process statistically rises the photon energy,
producing a distortion of the CMB black-body spectrum.  The final
result is a decrease (increase) of the CMB flux at frequencies smaller
(larger) than about 218 GHz.  The amplitude of this effect is directly
proportional to the Compton parameter $y(\theta)$, which is defined as
\begin{equation}\label{pres2}
y(\theta) = \frac{\sigma_{T}}{m_e c^2} \int P_e({\bf r}) \, dl\ ,
\end{equation}
\noindent 
where $\theta$ is the angular distance from the cluster centre,
$\sigma_T$ is the Thomson cross-section, and $P_e({\bf r}) \equiv
n_e({\bf r}) kT_e({\bf r})$ is the pressure of the ICM electrons at
the position {\bf r}; the integral is done along the line of sight.

The SZ effect can be expressed as a change in the brightness:
\begin{equation}
\Delta I_{\rm SZ} = g_{(x,T_e)}\ I_0 \ y\ , 
\label{eq:y}
\end{equation}
respectively; here $I_0 \equiv 2 (k T_{\rm CMB})^3 / (h c)^2$, $x
\equiv h \nu / k T_{\rm CMB}$, $T_{\rm CMB}$ is the present CMB
temperature and the function $g_{(x,T_e)}$ is given by:
\begin{equation}\label{gx}
g_{(x,T_e)} = \frac{x^4 e^x}{(e^x-1)^2} \left ( x \frac{e^x + 1}{e^x - 1} - 4
\right ) \left ( 1 + o(x, T_e) \right )\ ,
\label{eq:gx}
\end{equation}
and accounts for the frequency dependence of the SZ
effect; the term $o(x, T_e)$ represents the relativistic correction
\citep[see, e.g.,][]{1998ApJ...502....7I},
which, however, is negligible for clusters having $T \la 10$ keV.

We consider the Compton-$y$ parameter integrated over the entire
solid angle (and given in flux units) $y_{\Delta}$ defined as:
\begin{equation}\label{ydelta-m1}
y_{\Delta}=I_0 \! \!\int_0^{\theta_{\Delta}}\! \! \! y(\theta) d\Omega \ ;
\end{equation}
To remove the dependence of $y_{\Delta}$ on the angular 
diameter distance $d_{\rm a}(z)$ we use the intrinsic integrated Compton parameter $Y$, defined as:
\begin{equation}\label{ydelta-m1b}
Y \equiv d^2_{\rm a}(z)\, y_{\Delta}.
\end{equation}
The same quantity, but integrated over a fixed solid angle $\Omega$,
can be similarly written as:
\begin{equation}\label{ydelta-m12}
y_{\Omega} = I_0 \int_0^{\Omega} y(\theta) d\Omega\ .
\end{equation}
We fixed $\Omega=1$ arcmin, that
is $\la$ than the field of view of OVRO, used
in the observations of most of the sources in our sample \citep[see,
e.g.,][]{2003ApJ...591..515M}. Notice that in order to remove the
frequency dependence we have normalized $Y$, $y_{\Delta}$ and
$y_{\Omega}$ to $g_{(x,T_e)}$.

To integrate eqs. (\ref{ydelta-m1}) and (\ref{ydelta-m12}) we have recovered $y(\theta)$ from eq. \ref{pres2} by using the pressure profile $P({\bf q},P_0)$ determined in the spectral analysis (Sect. \ref{depr}), renormalized in such a way that $y(0)$ equals the central Comption parameter $y_0$ taken from the literature. This method can lead to systematics on $y_{\Omega}$ and $Y$ due to the fact that, even if we are assuming the true pressure profiles $P(r)$ in eq. (\ref{pres2}), $y_0$ has been obtained by assuming an isothermal $\beta$-model inferred from the brightness profile. The value of $y_0$ is thus potentially dependent on the underlying model of $P(r)$. As discussed in recent works \citep[see, e.g., ][]{2006ApJ...652..917L,2006ApJ...647...25B},
the relaxation of the isothermal assumption should apply to the
analysis of both X-ray and SZ data, to obtain a robust
and consistent description of the physics acting inside galaxy
clusters. Unfortunately, we have only the central Compton parameter, and not the complete $uv-$data, which are not public available: so it is very difficult to quantify the amplitude of this systematics, being $y_0$ determined through a best fit in the $uv-$plane.\\ 
Nevertheless, we can give an estimate in this way: we have computed the central Compton parameter $y^{I}_{\rm 0,X}$ inferred from the X-ray data by parametrizing first $P(r)$ in eq. (\ref{pres2}) with a $\beta$-model inferred on the brightness images: 
\begin{equation}\label{ydeltad}
y_0 = \frac{\sigma_{T}}{m_e c^2} \, n_0\, kT_{\rm gas}\int d\, x\, \left(1+ x^2\right)^{(1-3 \beta)/2}
\end{equation}
with $n_0=n_{\rm gas}(r=0)$ derived from the brightness profile $B(r)$:
\begin{equation}\label{eq:xsignal}
B(r=0) = \frac{1}{4\pi (1+z)^4} r_{\rm c} \Lambda\, 0.82\,n^2_0  \int d\,x \left(1+ x^2\right)^{1/2-3 \beta}
\end{equation}
where $\Lambda$ is the X-ray cooling function of the ICM in the
cluster rest frame in cgs units (erg cm$^3$ s$^{-1}$) integrated over the energy range of the brightness images ($0.5-5$ keV). Then we have calculated $y^{II}_{\rm 0,X}$ by accounting in eq. \ref{pres2} 
for the true pressure profile $P({\bf q},P_0)$ recovered by the spectral 
deprojection analysis (Sect. \ref{depr}), and therefore we determined the 
ratio $\eta=y^{II}_{\rm 0,X}/ y^{I}_{\rm 0,X}$. We notice that the parameter $\eta$ differs from the unity of $\lesssim 25$ per cent, comparable to statistical errors.

The different quantities related to the SZ effect 
are listed in Table \ref{tab:2} for all clusters in our sample.

\section{The X-ray and SZ scaling relations: theory and fitting procedure }\label{scales1}

\subsection{The scaling relations in the self-similar model}

The self-similar model \citep[see, e.g.,][]{1986MNRAS.222..323K} gives
a simple picture of the process of cluster formation in which the ICM
physics is driven by the infall of cosmic baryons into the
gravitational potential of the cluster DM halo.  The collapse and
subsequent shocks heat the ICM up to the virial temperature.  Thanks
to this model, which assumes that gravity is the only responsible for
the observed values of the different physical properties of galaxy
clusters, we have a simple way to establish theoretical analytic
relations between them.

Numerical simulations confirm that the DM component in clusters of
galaxies, which represents the dominant fraction of the mass, has a
remarkably self-similar behaviour; however the baryonic component does
not show the same level of self-similarity. This picture is confirmed
by X-ray observations, see for instance the deviation of the $L-T$
relation in clusters, which is steeper than the theoretical value
predicted by the previous scenario. These deviations from
self-similarity have been interpreted as the effects of
non-gravitational heating due to radiative cooling as well as the
energy injection from supernovae, AGN, star formation or galactic
winds \citep[see, e.g.,][]
{2001ApJ...546...63T,2001ApJ...555..597B,2002MNRAS.336..409B,
2002MNRAS.330..329B,2004MNRAS.348.1078B,2006ApJ...643..120B} which
make the gas less centrally concentrated and with a shallower profile
in the external regions with respect the DM component.  Consequently,
the comparison of the self-similar scaling relations to observations
allows us to evaluate the importance of the effects of the
non-gravitational processes on the ICM physics.

For $Y$ and $y_{\Omega}$ we have the following dependences on the cosmology:
\begin{equation}
{E_{z}}\Delta_z^{1/2} Y \propto 
\! {\left({{E_{z}}^{-1}\!\Delta_z^{-1/2}  y_0}\right)} 
  {\left({{E_{z}}\Delta_z^{1/2} R_{\Delta_z}}\right)}^{2}\ ,
\label{y0-y2}
\end{equation}
and
\begin{equation}
{E_{z}}^{-1}\!\Delta_z^{-1/2} y_{\Omega} \propto {E_{z}}^{-1}\!\Delta_z^{-1/2} y_0 \ ,
\end{equation}
respectively, where the factor $\Delta_z= 200
\times \left[1 +82 \left(\Omega_z-1\right) / \left(18 \pi^2\right) -39
\left(\Omega_z-1\right)^2 / \left(18 \pi^2\right) \right]$, with $\Omega_z = \Omega_{\rm 0m} (1+z)^3/E_z^2$, accounts for 
evolution of clusters in an adiabatic scenario \citep{1998ApJ...495...80B}.

Assuming the spherical collapse model for the DM halo
and the equation of hydrostatic equilibrium to describe the
distribution of baryons into the DM potential well, in the
self-similar model the cluster mass and temperature are related by:
\begin{equation}
E_{z} \Delta_z^{1/2} M_{\rm tot} \propto T^{3/2}\ ;
\end{equation}
so we have $ R_{\Delta_z} \propto
{\left({M/(\rho_{c,z}\Delta_z)}\right)}^{1/3} \propto
T^{1/2}E_{z}^{-1}\Delta_z^{-1/2}$.  By setting $f_z\equiv E_{z}
{(\Delta_z/\Delta)}^{1/2}$, from the previous equations we can easily
obtain the following relations \citep[see, e.g.,
][]{1998ApJ...504...27M, allen1998, 2004A&A...417...13E,
2005A&A...441..893A, 2005MNRAS.356.1477D, 2005ApJ...628..655V,
2005ApJ...633..781K}:

\begin{equation}
f_z \left(Y\right) \propto {\left(f_z^{-1}
y_0\right)}^{5/3}\ ,
\end{equation}

\begin{equation}
y_{\Omega}  \propto y_0\ ,
\end{equation}

\begin{equation}\label{112}
f_z^{-1}\, y_0 \propto T^{3/2}\ ,
\end{equation} 

\begin{equation}\label{114}
f_z^{-1}\, y_0  \propto f_z M_{\rm tot}\ , 
\end{equation}

\begin{equation}\label{113}
f_z^{-1}\, y_0 \propto {\left(f_z^{-1} L\right)}^{3/4}\ ,
\end{equation}

\begin{equation}
f_z \, Y \propto T^{5/2}\ ,
\end{equation}

\begin{equation}
f_z \, Y\propto {\left(f_z
M_{\rm tot}\right)}^{5/3}\ ,
\end{equation}

\begin{equation} 
f_z\, Y \propto {\left(f_z^{-1}L\right)}^{5/4}\ .
\end{equation}

We also remember here that for galaxy clusters similar scaling laws
exist also in the X-ray band \citep[see, e.g.,
][]{2004MNRAS.354..111E, 2005A&A...441..893A, 2005ApJ...633..781K,
2006ApJ...640..691V}:
\begin{equation}
f_z^{-1} L  \propto   T_{\rm gas}^2\ ,
\end{equation}

\begin{equation}
f_z M_{\rm tot}  \propto  T_{\rm gas}^{3/2}\ ,
\end{equation}

\begin{equation}
f_z^{-1} L \propto  (f_z M_{\rm tot})^{4/3}\ ,
\end{equation}

\begin{equation}
f_z M_{\rm gas} \propto  T_{\rm gas}^{3/2}\ ,
\end{equation}

\begin{equation}
f_z^{-1} L \propto {(f_z M_{\rm gas})}^{4/3}\ .
\end{equation}
In our work we have considered all the physical quantity at fixed overdensity ($\Delta_z=\Delta$), i.e. $f_z=E_{z}$ in the above equations.

\subsection{Fitting the scaling relations}

We describe here the method adopted to obtain the best-fitting
parameters in the scaling relations.  Since they are power-law
relations, we carry out a log-log fit:
\begin{equation}
\log(Y)=\alpha + A \log(X)\ ,
\end{equation}
where $X$ and $Y$ represent the independent and dependent variables,
respectively (hereafter $Y \!\! \mid \!\!X$); $\alpha$ and $A$ are the
two free parameters to be estimated.  However, in the considered
scaling relations it is unclear which variable should be considered as
(in)dependent.  Moreover both $X$- and $Y$-data have errors due to
measurement uncertainties, plus an intrinsic scatter. For these
reasons, the ordinary least squares (OLS) minimization approach is not
appropriate: in fact it does not take into account intrinsic scatter
in the data, and it is biased when errors affect the independent
variable.  So we decided to use the BCES (Bivariate Correlated Errors
and intrinsic Scatter) $(Y \!\! \mid
\!\!X)$ modification or the bisector modification BCES $(Y,X)$
proposed by \cite{1996ApJ...470..706A}, for which the best-fit results
correspond to the bisection of those obtained from minimizations in
the vertical and horizontal directions.  Both these methods are robust
estimators that take into account both any intrinsic scatter and the
presence of errors on both variables.

The results for the best-fit normalization $\alpha$ and slope $A$ for
the listed scaling relations are presented in Table~\ref{tab:fit},
where we also report the values of the total scatter
\begin{equation}
S=\left[ \sum_{j}
\left(\log Y_j -\alpha -A \log X_j \right)^2 / \nu \right]^{1/2}\ 
\label{eq:chi4}
\end{equation}
and of the intrinsic scatter $\hat{S}$ calculated as:
\begin{equation}
\hat{S} = \left[ \sum_{j}
\left(\left(\log Y_j -\alpha -A \log X_j \right)^2 
-\epsilon_{\log Y_j}^2\right) / \nu \right]^{1/2}\ ,
\label{eq:chi3}
\end{equation}
where $\epsilon_{\log Y_j} = \epsilon_{Y_j}/(Y_j \ln 10)$, with
$\epsilon_{Y_j}$ being the statistical error of the measurement $Y_j$,
and $\nu$ is the number of degrees of freedom ($\nu={N}-2$, with $N$ equal to total number of data).

Notice that in these fits the physical quantities ($L$, $M_{\rm tot}$,
$M_{\rm gas}$, $Y$) refer to $R_{2500}$ estimated through
the mass estimates based on the RTM model. 

\begin{table*}
\caption{
Best-fit parameters for the scaling relations computed by using the
cluster quantities evaluated at $R_{2500}$; masses are estimated using
the RTM profile.  For each relation we give the logarithmic slope $A$
(compared to the theoretically expected value $A^*$), the
normalization $\alpha$, the intrinsic scatter $\hat{S}$ and the
logarithmic scatter of the data $S$.  The results are given both for a
subsample including the CC clusters (11 objects), and for
the whole sample (24 objects). In the column ``method", symbols (1)
and (2) indicate if the fit has been performed by adopting the BCES
$(Y \!\mid \!X)$ or BCES $(Y,X)$ methods, respectively.  With the
notation ($y_{0,-4}$, $y_{\Omega}$, $Y_8$), 
$L_{44}$, $T_7$, $M_{14}$, we indicate the
Compton parameter, the X-ray luminosity, the temperature and the mass, in units
of ($10^{-4}$, mJy, $10^8$ mJy Mpc$^2$), $10^{44}$ erg s$^{-1}$, 7 keV, $10^{14}
M_{\odot}$, respectively. }
\begin{tabular}{l@{\hspace{.7em}} c@{\hspace{.7em}} c@{\hspace{.7em}} c@{\hspace{.7em}} | c@{\hspace{.7em}} c@{\hspace{.7em}} c@{\hspace{.7em}} c@{\hspace{.7em}} c@{\hspace{.7em}} c@{\hspace{.7em}}}
\hline \\
& \multicolumn{4}{c}{Cooling core clusters} & \multicolumn{4}{c}{All clusters} & \\
& \multicolumn{4}{c}{11 objects} & \multicolumn{4}{c}{24 objects} & \\
\hline \\
relation $(Y-X)$ & $A/A^*$ & $\alpha$ & $\hat{S}$ & $S$ & $A/A^*$ & $\alpha$ & $ \hat{S} $ & $S$ & method \\
\hline \\
$ f_z \,Y_8\! - \!{f_z^{-1} y_{0,-4}} $ &   1.22($ \pm 0.15 $)/1.67 & -1.07($ \pm 0.06 $) & 0.090 & 0.113 &    1.19($ \pm 0.20 $)/1.67 & -0.91($ \pm 0.06 $) & 0.116 & 0.137 & (1) \\
$ y_{\Omega} - y_{0,-4} $ &   0.93($ \pm 0.14 $)/1.00 & 0.61($ \pm 0.04 $) & 0.033 & 0.076 &    0.92($ \pm 0.26 $)/1.00 & 0.66($ \pm 0.08 $) & 0.120 & 0.140 & (1) \\ 
\hline \\
$ f_z^{-1} y_{0,-4} - T_{\rm ew,7}$ &   2.21($ \pm 0.32 $)/1.50 & 0.19($ \pm 0.05 $) & 0.138 & 0.154 &    2.06($ \pm 0.23 $)/1.50 & 0.15($ \pm 0.03 $) & 0.123 & 0.141 & (2) \\
$ f_z^{-1} \,  y_{0,-4} - f_z M_{\rm tot,14} $ &   1.25($ \pm 0.30 $)/1.00 & -0.50($ \pm 0.22 $) & 0.248 & 0.257 &    1.22($ \pm 0.29 $)/1.00 & -0.41($ \pm 0.17 $) & 0.211 & 0.222 & (2) \\
$ f_z^{-1}  \,  y_{0,-4} - f_z^{-1} L_{44} $ &   0.75($ \pm 0.07 $)/0.75 & -0.69($ \pm 0.11 $) & 0.156 & 0.170 &    0.61($ \pm 0.05 $)/0.75 & -0.48($ \pm 0.07 $) & 0.124 & 0.142 & (2) \\
$ f_z\, Y_8 - T_{\rm ew,7} $ &   2.74($ \pm 0.23 $)/2.50 & -0.83($ \pm 0.03 $) & 0.103 & 0.124 &    2.64($ \pm 0.28 $)/2.50 & -0.74($ \pm 0.03 $) & 0.139 & 0.157 & (2) \\
$ f_z \, Y_8 - f_z M_{\rm tot,14} $ &   1.56($ \pm 0.29 $)/1.67 & -1.70($ \pm 0.21 $) & 0.235 & 0.245 &    1.48($ \pm 0.39 $)/1.67 & -1.42($ \pm 0.21 $) & 0.288 & 0.297 & (2) \\
$ f_z \, Y_8- f_z^{-1}L_{44} $ &   0.92($ \pm 0.11 $)/1.25 & -1.92($ \pm 0.16 $) & 0.183 & 0.196 &    0.81($ \pm 0.07 $)/0.75 & -1.58($ \pm 0.10 $) & 0.237 & 0.248 & (2) \\
$ f_z^{-1}  y_{\Omega} - T_{\rm ew,7}$ &   1.98($ \pm 0.46 $)/1.50 & 0.80($ \pm 0.05 $) & 0.143 & 0.158 &    2.15($ \pm 0.45 $)/1.50 & 0.79($ \pm 0.05 $) & 0.167 & 0.182 & (2) \\
$ f_z^{-1}  y_{\Omega} - f_z M_{\rm tot,14} $ &   1.12($ \pm 0.31 $)/1.00 & 0.17($ \pm 0.22 $) & 0.239 & 0.249 &    1.07($ \pm 0.17 $)/1.00 & 0.31($ \pm 0.10 $) & 0.278 & 0.288 & (2) \\
$ f_z^{-1}  y_{\Omega}- f_z^{-1}L_{44} $ &   0.68($ \pm 0.09 $)/0.75 & -0.02($ \pm 0.14 $) & 0.160 & 0.174 &    0.74($ \pm 0.10 $)/0.75 & 0.00($ \pm 0.15 $) & 0.233 & 0.244 & (2) \\

\hline \\ 
$ f_z^{-1} L_{44} - T_{\rm ew,7} $ &   2.98($ \pm 0.53 $)/2.00 & 1.18($ \pm 0.05 $) & 0.182 & 0.183 &    3.37($ \pm 0.39 $)/2.00 & 1.03($ \pm 0.05 $) & 0.220 & 0.221 & (2) \\
$ f_z^{-1} L_{44} - f_z M_{\rm tot,14} $ &   1.71($ \pm 0.46 $)/1.33 & 0.24($ \pm 0.32 $) & 0.205 & 0.206 &    2.03($ \pm 0.54 $)/1.33 & 0.08($ \pm 0.32 $) & 0.269 & 0.270 & (2) \\
$ f_z M_{\rm tot,14} - T_{\rm ew,7} $ &   1.74($ \pm 0.25 $)/1.50 & 0.56($ \pm 0.03 $) & 0.000 & 0.098 &    1.69($ \pm 0.40 $)/1.50 & 0.47($ \pm 0.04 $) & 0.044 & 0.142 & (2) \\
$ f_z M_{\rm tot,14} - T_{\rm mw,7} $ &   1.63($ \pm 0.25 $)/1.50 & 0.54($ \pm 0.04 $) & 0.000 & 0.087 &    1.69($ \pm 0.34 $)/1.50 & 0.45($ \pm 0.03 $) & 0.000 & 0.125 & (2) \\
$ f_z M_{\rm gas,13} - T_{\rm ew,7} $ &   1.94($ \pm 0.21 $)/1.50 & 0.57($ \pm 0.02 $) & 0.083 & 0.086 &    2.09($ \pm 0.23 $)/1.50 & 0.51($ \pm 0.02 $) & 0.107 & 0.110 & (2) \\
$ f_z^{-1} L_{44} - f_z M_{\rm gas,13} $ &   1.55($ \pm 0.13 $)/1.33 & 0.30($ \pm 0.09 $) & 0.083 & 0.085 &    1.64($ \pm 0.13 $)/1.33 & 0.19($ \pm 0.09 $) & 0.131 & 0.132 & (2) \\

\hline \\ 
\end{tabular}

\label{tab:fit}
\end{table*}

\subsection{On the evolution of the scaling relations}

We can extend the previous analysis by investigating the redshift
evolution of the scaling relations at $z>0.1$.  Note that only two
objects are available at $z>0.6$ and that all CC clusters are at
redshift below 0.45.  We parametrize the evolution using a $(1+z)^B$
dependence and put constraints on the value of $B$ by considering a
least-square minimization of the relation
\begin{equation}
\log(Y)=\alpha + A \log(X) + B \log (1+z)\ .
\label{eq:bfit2}
\end{equation}
This is obtained by defining a grid of values of $B$ and looking for the
minimum of a $\chi^2$-like function, defined as:
\begin{equation}
\chi^2 = \sum_j \frac{\left[ \log Y_j -\alpha - 
A \log X_j -B \log (1+z_j) \right]^2}{\epsilon_{\log Y_j}^2
+\epsilon_{\alpha}^2 +A^2 \epsilon_{\log X_j}^2 +\epsilon_{A}^2
\log^2 X_j}\ ;
\label{eq:chi2}
\end{equation}
the sum is over all data, and $\epsilon_{\log X}\equiv\epsilon_{X}/(X
\ln 10)$ and $\epsilon_{\log Y}\equiv\epsilon_{Y}/(Y \ln 10)$ are
related to the uncertainties on $X$ and $Y$, respectively.  The
best-fit parameters values calculated by using this method are
reported in Table \ref{tab:fit_evol}.  Again in these fits, which
refer to same scaling relations presented in Table \ref{tab:fit}, the
physical quantities ($L$, $M_{\rm tot}$, $M_{\rm gas}$,
$Y$) refer to $R_{2500}$, and masses are computed by
assuming the RTM model.

\begin{table*}
\caption{
Best-fit parameters for the redshift evolution of the scaling
relations.  Again, the quantities are evaluated at $R_{2500}$ and
masses are estimated by using the RTM profile.  For each relation we
list the redshift evolution parameter $B$, the logarithmic slope $A$
(compared to the theoretically expected value $A^*$), the
normalization $\alpha$, the minimum value of the function $\chi^2$ and
the number of degrees of freedom (d.o.f.).  The results are given both
for a subsample including the CC-only clusters (11 objects), and for the whole
sample (24 objects).  With the
notation ($y_{0,-4}$, $y_{\Omega}$, $Y_8$), 
$L_{44}$, $T_7$, $M_{14}$, we indicate the
Compton parameter, the X-ray luminosity, the temperature and the mass, in units
of ($10^{-4}$, mJy, $10^8$ mJy Mpc$^2$), $10^{44}$ erg s$^{-1}$, 7 keV, $10^{14}
M_{\odot}$, respectively. }
\begin{tabular}{l@{\hspace{.7em}} c@{\hspace{.7em}} c@{\hspace{.7em}} c@{\hspace{.7em}} | c@{\hspace{.7em}} c@{\hspace{.7em}} c@{\hspace{.7em}} c@{\hspace{.7em}} c@{\hspace{.7em}} c@{\hspace{.7em}} }
\hline \\
& \multicolumn{4}{c}{Cooling core clusters} & & \multicolumn{4}{c}{All clusters} \\
& \multicolumn{4}{c}{11 objects} && \multicolumn{4}{c}{24 objects} \\
\hline \\ 
relation $(Y-X)$ & B & $A/A^*$ & $\alpha$ &  $ {\chi}^2_{\rm min}$ (d.o.f.) &  & B & $A/A^*$ & $\alpha$ &  $ {\chi}^2_{\rm min}$ (d.o.f.)\\ 
\hline \\
$ f_z \, Y_8\! - \!{f_z^{-1} y_{0,-4}} $ &   2.36$_{- 0.68 }^{+0.64}$  & 1.02($ \pm 0.09)/1.67$ & -1.24($ \pm 0.04 $)  & 15.9(8) &  &    0.76$_{- 0.28 }^{+0.28}$  & 1.15($ \pm 0.08)/1.67$ & -1.00($ \pm 0.03 $)  & 87.8(21) \\
$ y_{\Omega} - y_{0,-4} $ &   -1.56$_{- 0.60 }^{+0.56}$  & 0.85($ \pm 0.08)/1.00$ & 0.80($ \pm 0.04 $)  & 5.1(8) &  &    -1.24$_{- 0.24 }^{+0.24}$  & 0.82($ \pm 0.07)/1.00$ & 0.83($ \pm 0.03 $)  & 88.6(21) \\

\hline \\
$ f_z^{-1} y_{0,-4} - T_{\rm ew,7} $ &   -2.12$_{- 0.96 }^{+0.96}$  & 2.41($ \pm 0.25)/1.50$ & 0.40($ \pm 0.03 $)  & 23.4(8) &  &    0.08$_{- 0.32 }^{+0.36}$  & 2.08($ \pm 0.17)/1.50$ & 0.12($ \pm 0.02 $)  & 82.4(21) \\
$ f_z^{-1} \,  y_{0,-4} - f_z M_{\rm tot,14} $ &   -2.44$_{- 2.52 }^{+1.68}$  & 1.35($ \pm 0.23)/1.00$ & -0.33($ \pm 0.19 $)  & 29.6(8) &  &    0.08$_{- 0.48 }^{+0.48}$  & 0.98($ \pm 0.10)/1.00$ & -0.27($ \pm 0.07 $)  & 55.6(21) \\
$ f_z^{-1}  \,  y_{0,-4} - f_z^{-1} L_{44} $ &   0.04$_{- 0.48 }^{+0.48}$  & 0.69($ \pm 0.05)/0.75$ & -0.57($ \pm 0.07 $)  & 48.0(8) &  &    -0.32$_{- 0.16 }^{+0.16}$  & 0.62($ \pm 0.03)/0.75$ & -0.44($ \pm 0.05 $)  & 99.7(21) \\
$ f_z\, Y_8 - T_{\rm ew,7} $ &   -1.08$_{- 1.16 }^{+1.12}$  & 2.98($ \pm 0.31)/2.50$ & -0.71($ \pm 0.04 $)  & 10.6(8) &  &    0.28$_{- 0.40 }^{+0.40}$  & 2.66($ \pm 0.20)/2.50$ & -0.78($ \pm 0.03 $)  & 37.9(21)  \\
$ f_z \, Y_8 - f_z M_{\rm tot,14} $ &   -2.32$_{- 2.60 }^{+1.96}$  & 1.68($ \pm 0.25)/1.67$ & -1.55($ \pm 0.20 $)  & 20.5(8) &  &    0.28$_{- 0.52 }^{+0.52}$  & 1.14($ \pm 0.13)/1.67$ & -1.26($ \pm 0.08 $)  & 78.9(21)\\
$ f_z \, Y_8- f_z^{-1}L_{44} $ &   2.40$_{- 0.48 }^{+0.44}$  & 0.70($ \pm 0.05)/1.25$ & -1.81($ \pm 0.07 $)  & 58.9(8) &  &    0.12$_{- 0.16 }^{+0.20}$  & 0.62($ \pm 0.04)/0.75$ & -1.35($ \pm 0.05 $)  & 206.0(21)  \\
$ f_z^{-1} y_{\Omega} - T_{\rm ew,7} $ &   -4.00$_{- 0.96 }^{+0.96}$  & 2.31($ \pm 0.26)/1.50$ & 1.20($ \pm 0.03 $)  & 16.6(8) &  &    -1.44$_{- 0.32 }^{+0.36}$  & 1.95($ \pm 0.18)/1.50$ & 0.97($ \pm 0.02 $)  & 71.2(21) \\
$ f_z^{-1} y_{\Omega} - f_z M_{\rm tot,14} $ &   -4.88$_{- 2.44 }^{+1.68}$  & 1.33($ \pm 0.22)/1.00$ & 0.52($ \pm 0.18 $)  & 26.9(8) &  &    -1.52$_{- 0.40 }^{+0.44}$  & 0.82($ \pm 0.12)/1.00$ & 0.64($ \pm 0.08 $)  & 120.0(21) \\
$ f_z^{-1} y_{\Omega}- f_z^{-1}L_{44} $ &   -1.52$_{- 0.44 }^{+0.48}$  & 0.58($ \pm 0.05)/0.75$ & 0.34($ \pm 0.07 $)  & 47.5(8) &  &    -1.72$_{- 0.16 }^{+0.16}$  & 0.40($ \pm 0.03)/0.75$ & 0.65($ \pm 0.05 $)  & 178.0(21) \\

\hline \\  
$ f_z^{-1} L_{44} - T_{\rm ew,7} $ &   -0.72$_{- 1.00 }^{+0.96}$  & 3.27($ \pm 0.29)/2.00$ & 1.25($ \pm 0.03 $)  & 69.40(8) &  &    0.92$_{- 0.52 }^{+0.52}$  & 4.05($ \pm 0.24)/2.00$ & 0.85($ \pm 0.03 $)  & 190.0(21)  \\
$ f_z^{-1} L_{44} - f_z M_{\rm tot,14} $ &   -1.32$_{- 1.56 }^{+1.24}$  & 1.29($ \pm 0.16)/1.33$ & 0.66($ \pm 0.13 $)  & 5.4(8) &  &    -0.24$_{- 0.56 }^{+0.56}$  & 1.36($ \pm 0.11)/1.33$ & 0.53($ \pm 0.07 $)  & 40.5(21)\\
$ f_z M_{\rm tot,14} - T_{\rm ew,7} $ &   0.56$_{- 1.20 }^{+1.12}$  & 1.79($ \pm 0.30)/1.50$ & -1.00($ \pm 0.28 $)  & 11.1(8) &  &    -0.08$_{- 0.52 }^{+0.52}$  & 2.30($ \pm 0.24)/1.50$ & -1.51($ \pm 0.22 $)  & 48.4(21) \\
$ f_z M_{\rm tot,14} - T_{\rm mw,7} $ &   -0.88$_{- 1.32 }^{+1.24}$  & 2.00($ \pm 0.28)/1.50$ & -1.09($ \pm 0.27 $)  & 7.1(8) &  &    -0.32$_{- 0.48 }^{+0.48}$  & 2.32($ \pm 0.22)/1.50$ & -1.54($ \pm 0.21 $)  & 33.0(21) \\
$ f_z M_{\rm gas,13} - T_{\rm ew,7} $ &   0.16$_{- 0.60 }^{+0.56}$  & 2.00($ \pm 0.16)/1.50$ & 0.57($ \pm 0.02 $)  & 39.2(8) &  &    0.84$_{- 0.28 }^{+0.28}$  & 2.17($ \pm 0.12)/1.50$ & 0.41($ \pm 0.02 $)  & 127.0(21)\\
$ f_z^{-1} L_{44} - f_z M_{\rm gas,13} $ &   -0.92$_{- 0.24 }^{+0.24}$  & 1.43($ \pm 0.03)/1.33$ & 0.45($ \pm 0.03 $)  & 73.7(8) &  &    -0.60$_{- 0.12 }^{+0.12}$  & 1.63($ \pm 0.02)/1.33$ & 0.26($ \pm 0.02 $)  & 358.0(21) \\

\hline \\
\end{tabular}

\label{tab:fit_evol}
\end{table*}

\section{Discussion of the results}\label{cal}

We present here a general discussion of our results concerning the
scaling relations. In particular we have chosen to consider both the
whole sample (CC plus NCC objects) and the CC-only subsample:
this is done to allow a more direct comparison of our results with
most of the works present in the literature, which are based on
CC-sources only.  Moreover this allows also to obtain at the same
time more general relations which can be useful for future much extended
(X-ray and SZ) cluster surveys, in which the distinction between
relaxed and unrelaxed systems will be not easy.

\subsection{The X-ray-only scaling relations}

In this section we consider the scaling relations involving quantities
extracted from the X-ray data only.  We start by examining the
relation between $M_{\rm tot}$ and $T$, and finding in general a good agreement between our
best-fitting slopes and the values expected in the self-similar model. Then we will consider the other X-ray relations, finding slopes which
are steeper than expected from the self-similar model.  In particular
the $L-T$, $L-M_{\rm gas}$ and $M_{\rm gas}-T$ relations display
deviations larger than $2\sigma$, while for the $L-M_{\rm tot}$
relation we found agreement between the observed slope and the expected one.

\subsubsection{The $M_{\rm tot}-T$ relation}\label{M-Tmwrela1}

Without any assumption for models on 
the gas density and (deprojected)
temperature profile, we have supposed that the DM density profile is
well described by an analytical model (RTM or NFW).  
Thanks to the results of numerical simulations, we know,
indeed, sufficiently well the DM physics which is in fact very simple,
only depending on the gravity, unlike the physics of the baryons,
which is also affected by further sources of non-gravitational energy.
Moreover we have removed the observational biases in the determination
of the deprojected temperature (and consequently of the mass) by
adopting the spectral-like temperature estimator (see
Sect. \ref{depr}).  In this way we have a bias-free 
estimate of the deprojected temperature and,
therefore, of the cluster mass.  Below we focus our attention on
$T_{\rm mw}$, because it is directly related to the total energy of
the particles and so comparable to the results of hydrodynamical
simulations, unlike $T_{\rm ew}$, which is affected by observational
biases \citep[see, e.g., ][]{2004MNRAS.351..505G,2004MNRAS.354...10M, 2001ApJ...546..100M}.

First, we notice that the two different models for the DM profile give
slightly different results. Nevertheless, at 
an overdensity of $\Delta=2500$ the
masses determined by using RTM are in perfect agreement with the
ones determined by using NFW ($\alpha^{\rm CC}_{\rm RTM}=0.540\pm 0.037$ and $A^{\rm CC}_{\rm RTM}=1.630\pm 0.253$, $\alpha^{\rm CC}_{\rm NFW}=0.546\pm 0.035$ and $A^{\rm CC}_{\rm NFW}=1.590\pm 0.250$). At $\Delta=500$ the situation becomes
less clear, because for most of the clusters we needed to extrapolate
from $R_{\rm spec}$ (corresponding to $\Delta\sim 1000$) up to $\Delta=500$, being $R_{\rm spec}$ of order of $(1/3)$-$(1/2)$ of the 
virial radius (roughly corresponding to $R_{2500}-R_{1000}$). Hereafter we consider only the RTM model, even if most of the results present 
in the literature are usually based on the NFW one.  

Considering the whole
sample, we find a normalization ($\alpha= 0.45 \pm 0.03$), which is
$\sim 10$ ($\sim 5$) per cent smaller than the value found by \cite{2001MNRAS.328L..37A} (\cite{2005A&A...441..893A}), who only consider relaxed clusters.  Our
normalization ($\alpha= 0.54 \pm 0.04$) is instead $\sim 10$ ($\sim
15$) per cent larger than the value of Allen et al. (Arnaud et al.) if
we only consider the CC-only subsample. This suggests a different
behaviour depending on the presence or not of a cooling core (see also
the left panel Fig. \ref{L-T3rela}): 
in fact we find that at $\Delta = 2500$ the
normalization of the NCC subsample at $M_{2500}=5\times 10^{14}M_{\odot}$ (corresponding to our median value for the mass) 
is $\approx 10$ per cent smaller than
for the CC-only subsample; conversely at $\Delta = 500$ the two
subsamples give consistent normalizations, but the robustness of this
result is affected by the fact that in this case we have to
extrapolate the mass profile out of the region covered by
observational data.

Some other authors \citep[e.g.,][]{2005A&A...441..893A} prefer to mask
out the central region (up to 0.1$\times R_{200}$) in the
determination of the mass profile. We have decided to check the
effects of the inclusion of the cooling region in our analysis by
comparing the values of the mass obtained by excluding or not the
central 100 kpc in the determination of the best fit parameters of the
RTM profile: we pointed out that accounting for the cooling region
does not involve any systematic error on the determination of the
mass, indeed we obtain more statistically robust results.

Consequently the disagreement between CC and NCC clusters is probably
due to a different state of relaxation, namely that the former are
more regular and with more uniform physical properties than the latter
\citep{2002ApJ...567..163D}; this is true even if we have masked out
the most evident substructures.  Notice that the observed mismatch is
only marginally statistically significant ($\sim 1-1.5\sigma$).  For a
couple of clusters, namely A520 and A2163, we find that the exclusion
of the unrelaxed central regions avoids observational biases due to
the presence of local substructures: in particular the mass of the
first (second) object increases by a factor of $\sim 2$ ($\sim 1.5$)
when excluding the central 300 (360) kpc.  For other clusters which
are evidently unrelaxed, we did not find any convenient way to avoid
possible biases: even after masking out the most visible
substructures, the analysis of the density and deprojected temperature
profiles still reveals the possible presence of local irregularities
(a sort of local `jumps' in the profiles), which are difficult to
individuate in the brightness image.

At $\Delta=2500$, the best fitting normalization obtained considering
the whole sample is $\sim 30$ per cent below the value found in the
non-radiative hydrodynamic simulations by \cite{2001ApJ...546..100M}
\footnote{We have rescaled 
their results from $\Delta=500$ to $\Delta=2500$.}; for the CC-only
subsample, the normalization is $\sim 20$ per cent below the
theoretical value. The discrepancy is slightly reduced ($\sim 15-20$
per cent) with respect to the adiabatic hydrodynamic simulations by
\cite{1996ApJ...469..494E}.

The picture emerging from numerical simulations with a more
sophisticated ICM modeling is different.  The simulation by
\cite{2004MNRAS.348.1078B}, which includes radiative processes, supernova
feedback, galactic winds and star formation, 
suggests a normalization which is in
rough agreement with our whole sample, and 15 per cent lower with
respect to the CC-only  subsample.  Notice, however, that the
re-analysis of the same simulation data made by
\cite{2005ApJ...618L...1R}, who adopted a different definition of
temperature, the spectroscopic-like one (which is not consistent with
our definition of mass-weighted temperature; see above for a more
detailed discussion), gives a higher ($\sim40- 50$ per cent)
normalization.

Finally we notice that the slope of the $M-T$ relation is, indeed, in
agreement with the theoretical expectations ($A^*=1.5$).

Considering the results at an overdensity of $500$, we found a good
agreement (at 1$\sigma$ level) between observed and theoretical
slopes.

Our analysis suggests no evolution ($B^{\rm CC}=-0.88^{+1.24}_{-1.32}$, $B^{\rm all}=-0.32\pm 0.48$), in agreement with the literature 
\citep[see, e.g.,][]{2001A&A...368..749F,2004A&A...417...13E,2001MNRAS.328L..37A}. 

We compare also our intrinsic scatter, which is consistent with 
zero, with the one estimated by \cite{2005ApJ...618L...1R}:
they find a scatter of $\approx 30 (16)$ per cent by considering the
emission-weighted (spectroscopic-like) temperature.  We reach
similar conclusions comparing our intrinsic scatter with the value
retrieved by \cite{2005ApJ...623L..63M}.

\subsubsection{The $L-T$ relation}\label{lll33}

\begin{figure*}
\hbox{
\psfig{figure=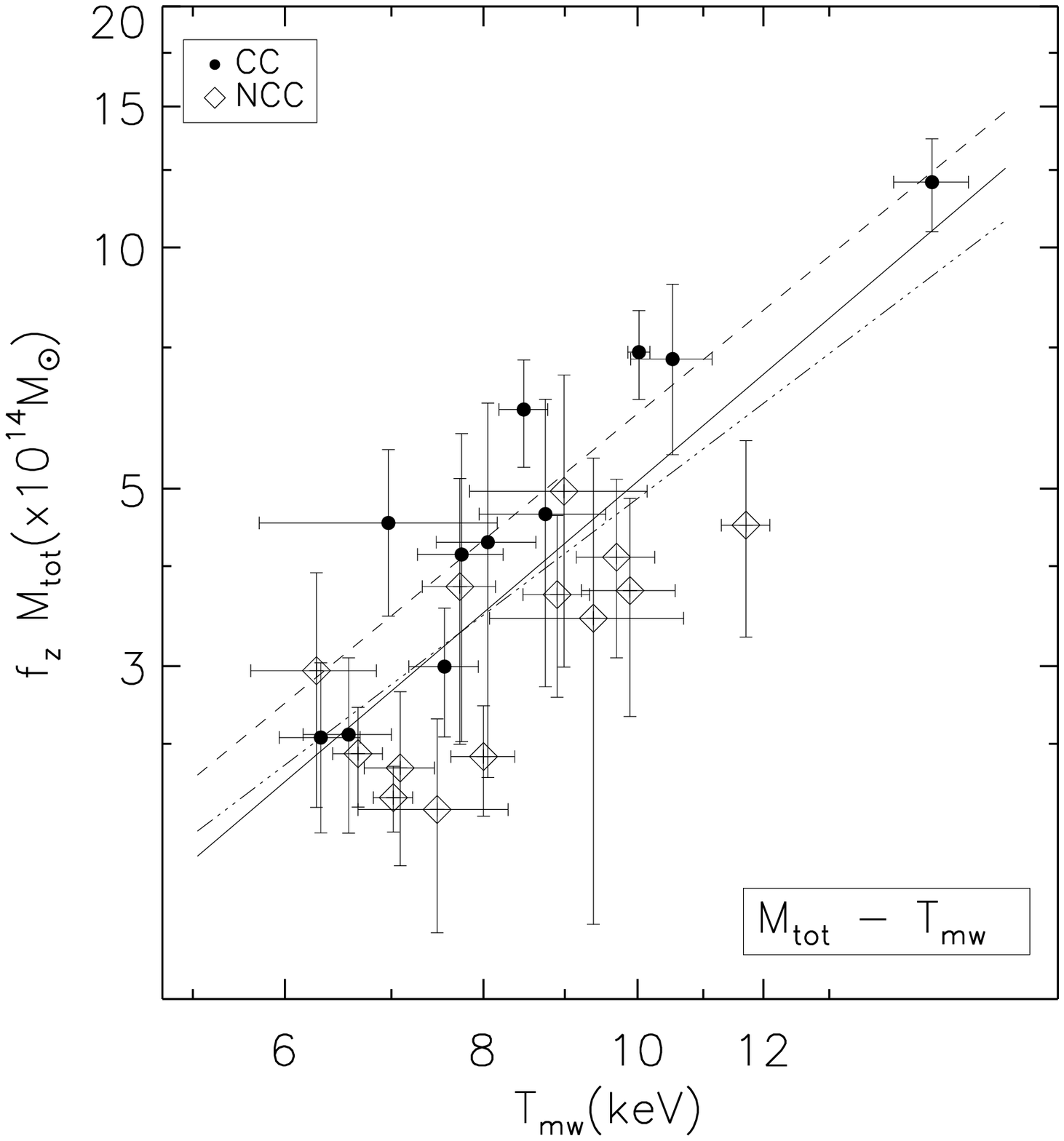  ,width=0.44\textwidth}
\psfig{figure=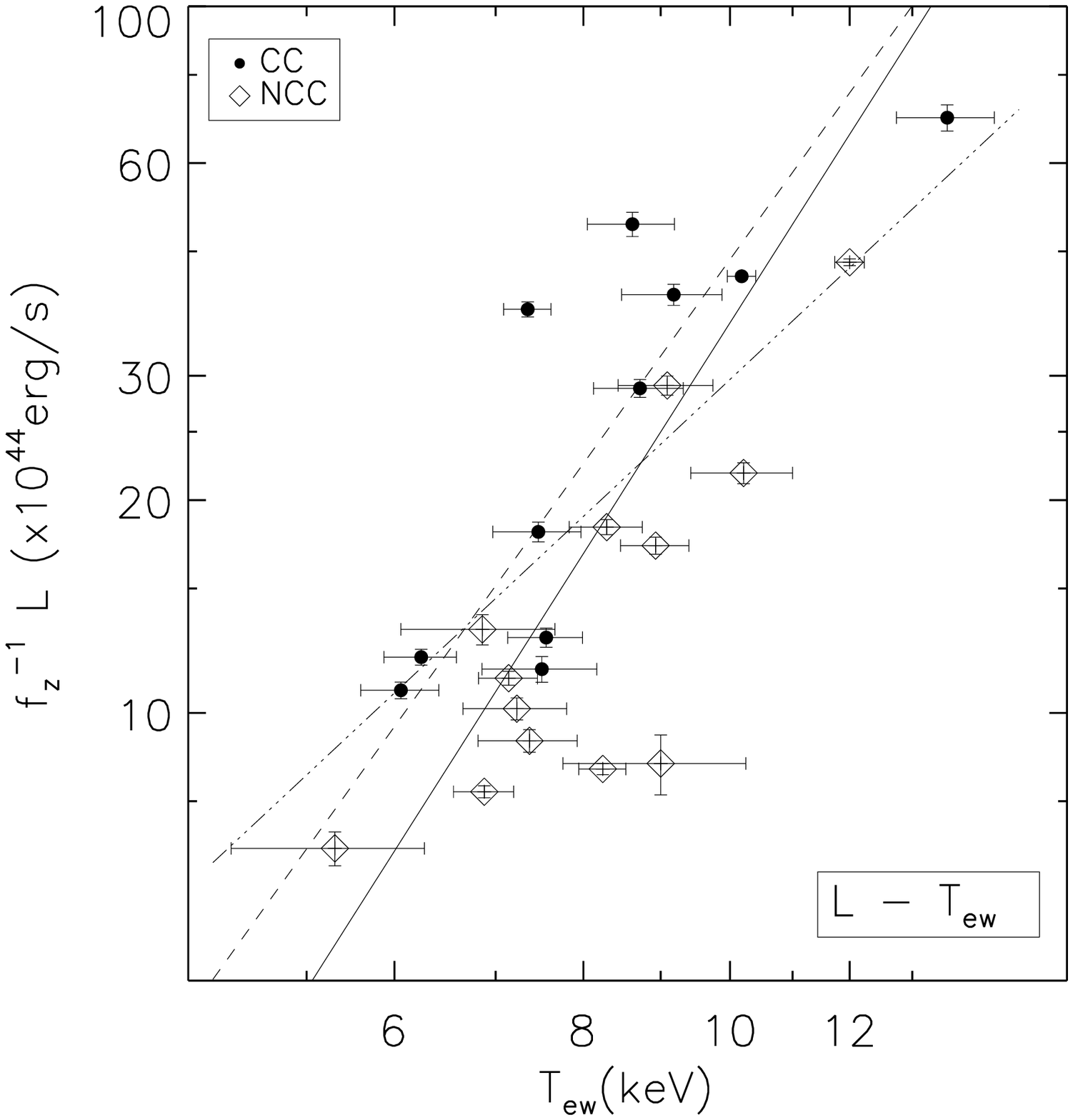  ,width=0.44\textwidth}
}
\caption[]{
The relations between $M_{\rm tot}$-$T_{\rm mw}$ (left panel) and
$L$-$T_{\rm ew}$ (right panel).  In each panel the filled
circles represent cooling core (CC) sources, while the diamonds
are the no-cooling core (NCC) ones. The solid line refers to the best-fit
relation obtained when considering all clusters of our sample,
the dashed one represents the best-fit when the CC sources only have
been considered and the dot-dashed is the best-fit obtained by fixing the slope to the self-similar value.}
\label{L-T3rela} 
\end{figure*}

We find (see the upper-right panel of Fig. \ref{L-T3rela}) a marginal agreement of our results on the slope of this relation
($A^{\rm all}=3.37\pm 0.39$), with those obtained by \cite{2002A&A...391..841E},
who found $A=2.64\pm 0.64$ at $\Delta=2500$: however, 
their sample contains colder objects, for which a flatter
relation would be expected.  Our results also agree with the analysis
made by \cite{1998ApJ...504...27M}: $A=2.64\pm 0.27$. 
Notice that his cluster sample is not directly
comparable with ours, since it covers different ranges in redshift and
temperature.

We compare our results about the scatter ($\hat{S}=0.220$ and 
$S=0.221$) with those obtained by
\cite{1998ApJ...504...27M}, who found a smaller value: $S=0.103$ (see, however,
the previous comments on the different characteristics of the two
samples).

Moreover, we find (at $\sim 1\sigma$) a positive (negative) redshift 
evolution for all clusters (CC-only subsample), i.e. we notice a mildly 
different behaviour on the evolution CC and NCC clusters. For comparison  
\cite{2002A&A...391..841E} found $B=-1.04 \pm
0.32$ for their sample of clusters at higher redshift.\\

Regarding the normalization we observe a slightly different behaviour when the CC-only subsample and whole sample are considered: $\alpha^{\rm CC}=1.18\pm0.05$ and $\alpha^{\rm all}=1.03\pm0.05$, respectively. We notice that the luminosity of the CC clusters is systematically larger than that of the NCC clusters, even if we have corrected it for the cooling flow (see Sect. \ref{phy}), as already observed by \cite{1994ARA&A..32..277F}. On the contrary numerical simulations predict that the removal of the gas from the X-ray emitting phase reduces the luminosity \citep{2002MNRAS.336..527M}. This confirms that cooling \citep[][]{2000ApJ...544L...1B, 2001Natur.414..425V}  is not effective in removing baryons from the X-ray phase, because of the presence of an extra-source of feedback or pre-heating \citep[][]{1999MNRAS.307..463B,1998ApJ...501..493C, 2001ApJ...546...63T, 2002MNRAS.330..329B}, which maintains the ICM at warm temperature \citep{2002MNRAS.336..409B}. Alternatively, the more evident negative evolution of the CC clusters compared to the NCC ones (especially in the $y_{\Omega}-$X-ray(SZ) relations) could indicate different states of relaxation, being the former more regular, relaxed and virialized than the latter \citep{2002ApJ...567..163D}.

\subsubsection{Other X-ray scaling relations}
Here we discuss our results for the relations not shown in the figures. For the $M_{\rm gas}-T$ relation we find a $\sim
1\sigma$ discrepancy between the slope of this relation in the CC-only
subsample ($A^{\rm CC}=1.94\pm 0.21$ and $A^{\rm all}=2.09\pm 0.23$) and the theoretical expectation for the
self-similar model ($A=1.5$).  Nevertheless, our estimate is
consistent with the results already present in the literature.  By
applying a $\beta$-model to recover the gas mass,
\cite{1998ApJ...502..558V} measured $A=1.71\pm 0.23$ at the baryon
overdensity 1000 (approximately corresponding to the virial DM overdensity). 
Our slope is also in good agreement with the
value ($A=1.98\pm 0.18$) found by \cite{1999ApJ...517..627M}, always
by applying the $\beta$-model.  We have also a marginal
agreement (at $1\sigma$ level) with the value found by
\cite{2004A&A...417...13E} ($A=2.37\pm 0.24$), who make use of the
$\beta$-model and apply the correction for $E_z$.  Finally
\cite{2002A&A...391..841E}, combining a spectral analysis and the
application of a $\beta$-model to the brightness distribution and
without correcting for $E_z$, found $A=1.91\pm 0.29$ for $\Delta=2500$
and $A=1.74\pm 0.22$ at $\Delta=500$.  The results of this last paper
also suggest a low intrinsic scatter, in good agreement with our
analysis ($\hat{S}=0.079$). We point out here that we find some differences between CC and NCC
clusters at $\Delta=2500$, because of the contribution of the cooling
core region ($\lesssim 100$ kpc); at $\Delta=500$ this effect becomes
negligible because the behaviour of the gas mass is dominated by the
contribution from the external regions ($M_{\rm gas} \propto r$). Finally no significant evolution is observed ($B=0.16^{+0.56}_{-0.60}$) for the CC clusters; when we consider the whole sample, we notice a more significant positive evolution ($B=0.84 \pm 0.28$)

Regarding the $L-M_{\rm tot}$ the best-fit slope for the CC-only subsample
($A^{\rm CC}=1.71\pm0.46$) is in good agreement with the results obtained by
\cite{2002ApJ...567..716R} ($A=1.80\pm 0.08$),
\cite{2002A&A...391..841E} ($A=1.84\pm 0.23$) and
\cite{2004A&A...417...13E} ($A=1.88\pm 0.42$).
The observed scatter we measure ($S^{\rm CC}= 0.206$, $S^{\rm all}= 0.270$) is slightly smaller than
in previous analysis by \cite{2002ApJ...567..716R} ($S=0.32$), and in
agreement with \cite{2002A&A...391..841E} ($S=0.26$). This seems to
suggest that the methods we applied to correct the observed luminosity
(see Sect.~\ref{phy}) and to determine the total mass are quite
robust. Hints of negative evolution are observed ($B^{\rm CC}=-1.32^{+1.24}_{-1.56}$, $B^{\rm all}=-0.24\pm 0.56$).

For the $L-M_{\rm gas}$ law we measure a slope which is discrepant with respect to the theoretical value expected in the self-similar
model.  This relation, together with the one between $M_{\rm gas}-T$ and 
$M_{\rm tot}-T$, has the lowest intrinsic scatter 
between the X-ray only scaling laws.
Moreover we have a significant evidence of a
negative redshift evolution.

\subsection{The scaling relations involving the SZ effect}

\begin{figure*}
\hbox{
\psfig{figure=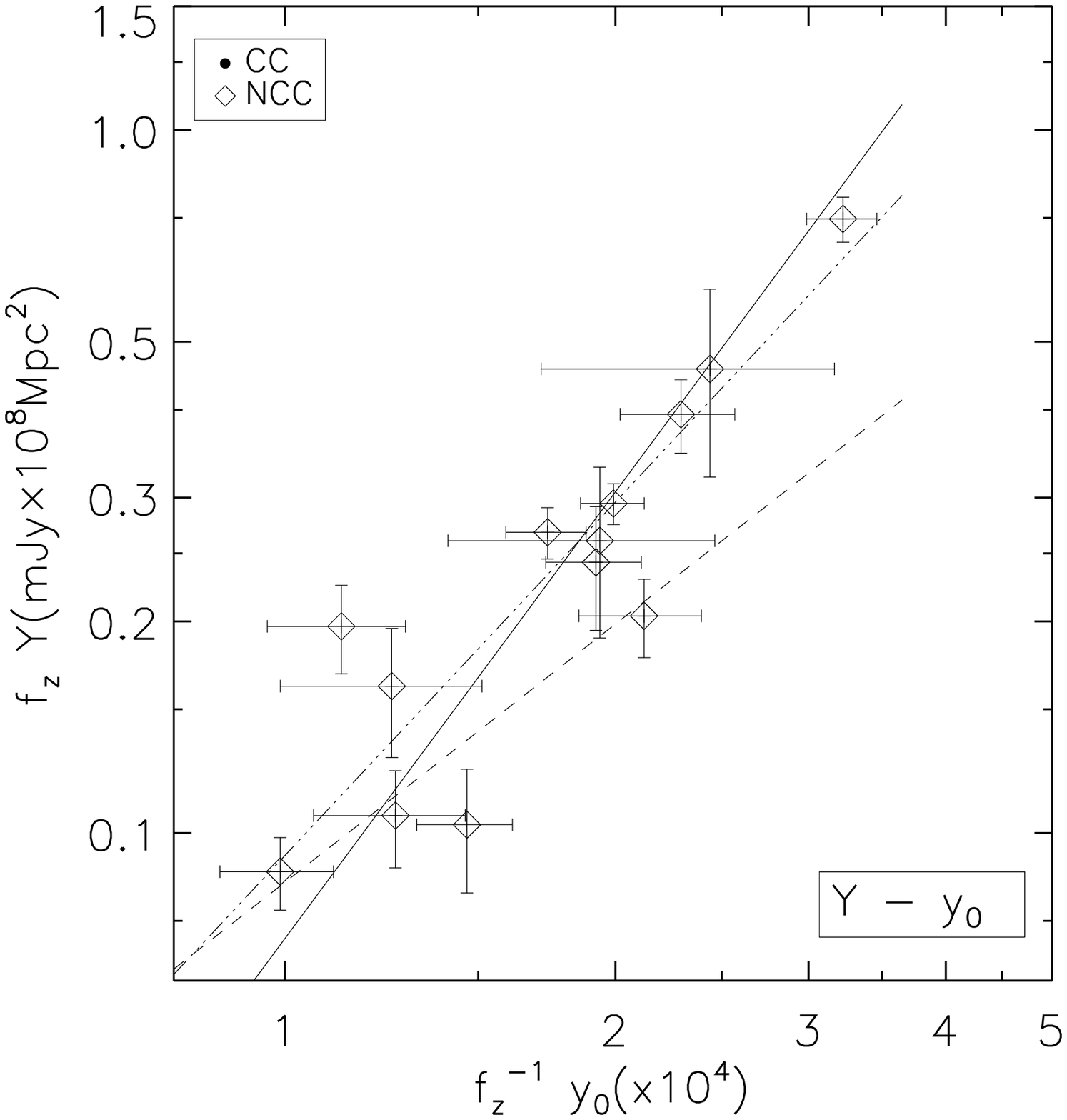,width=0.45\textwidth}
\psfig{figure=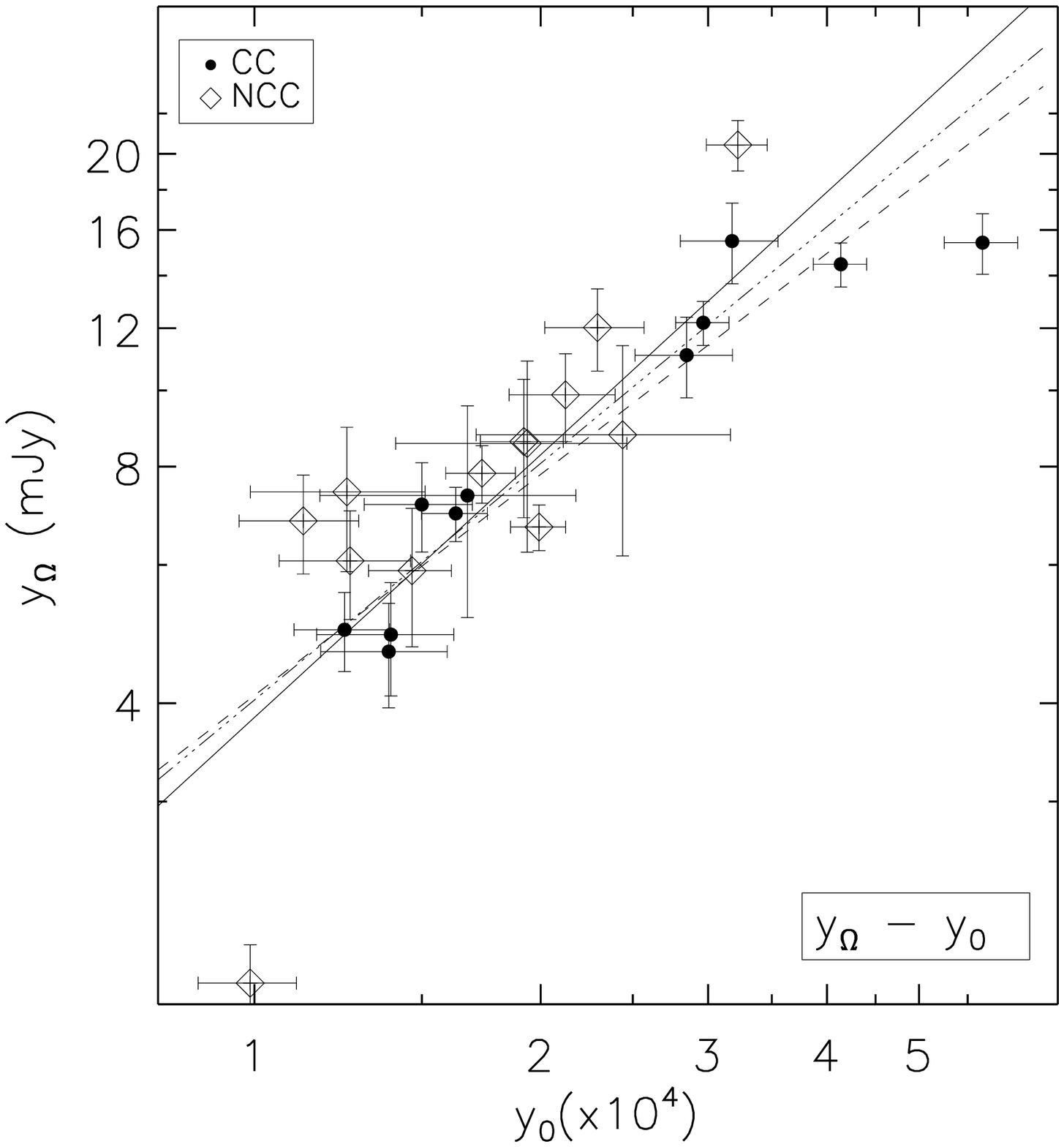,width=0.45\textwidth}
}
\caption[]{
As in Fig. \ref{L-T3rela} but for the relations between $Y-y_0$ (left panel) and
$y_{\Omega}-y_0$ (right panel).}
\label{y_0-relaSZ} 
\end{figure*}

\begin{figure*}
\hbox{
\psfig{figure=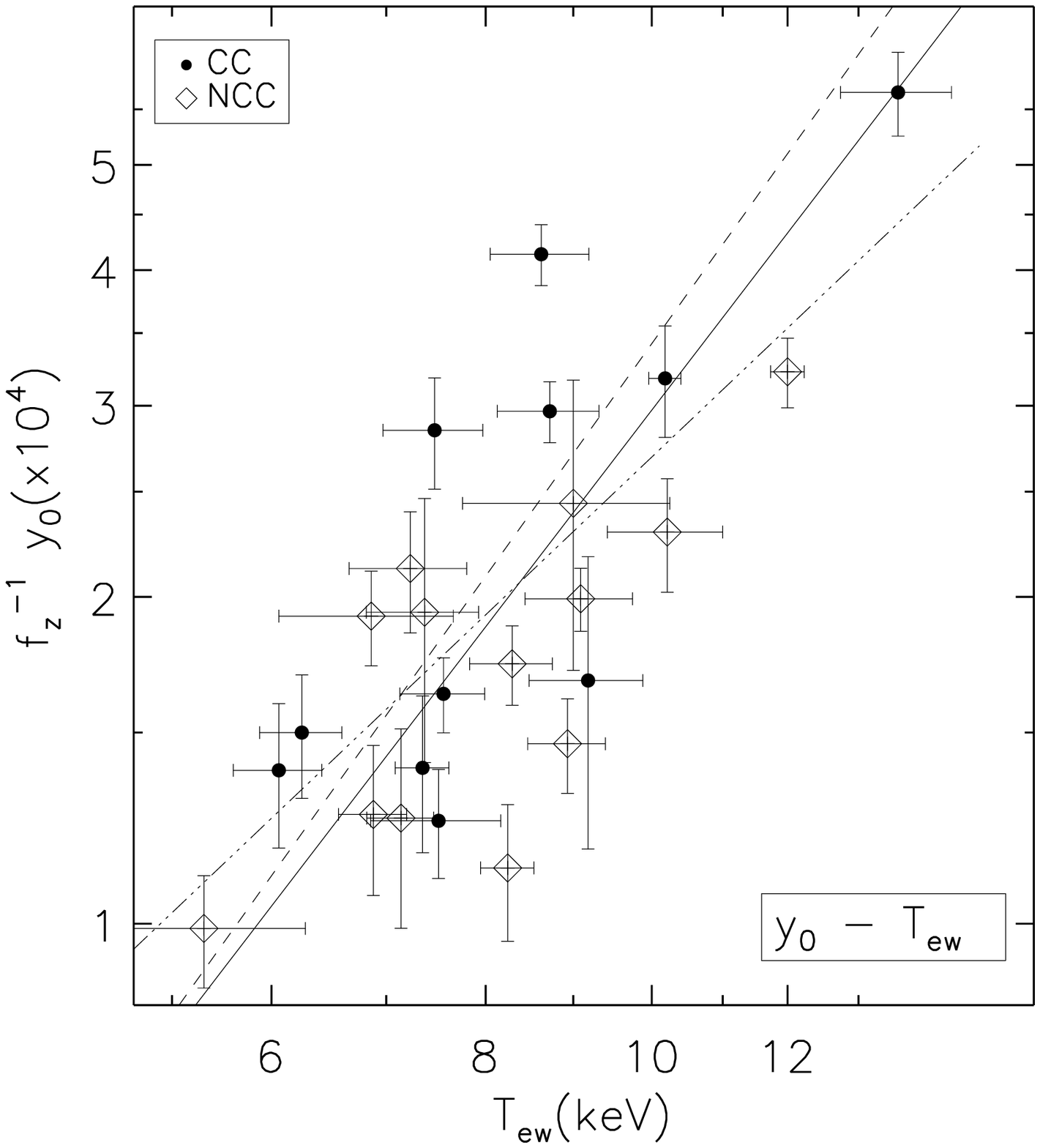,width=0.33\textwidth}
\psfig{figure=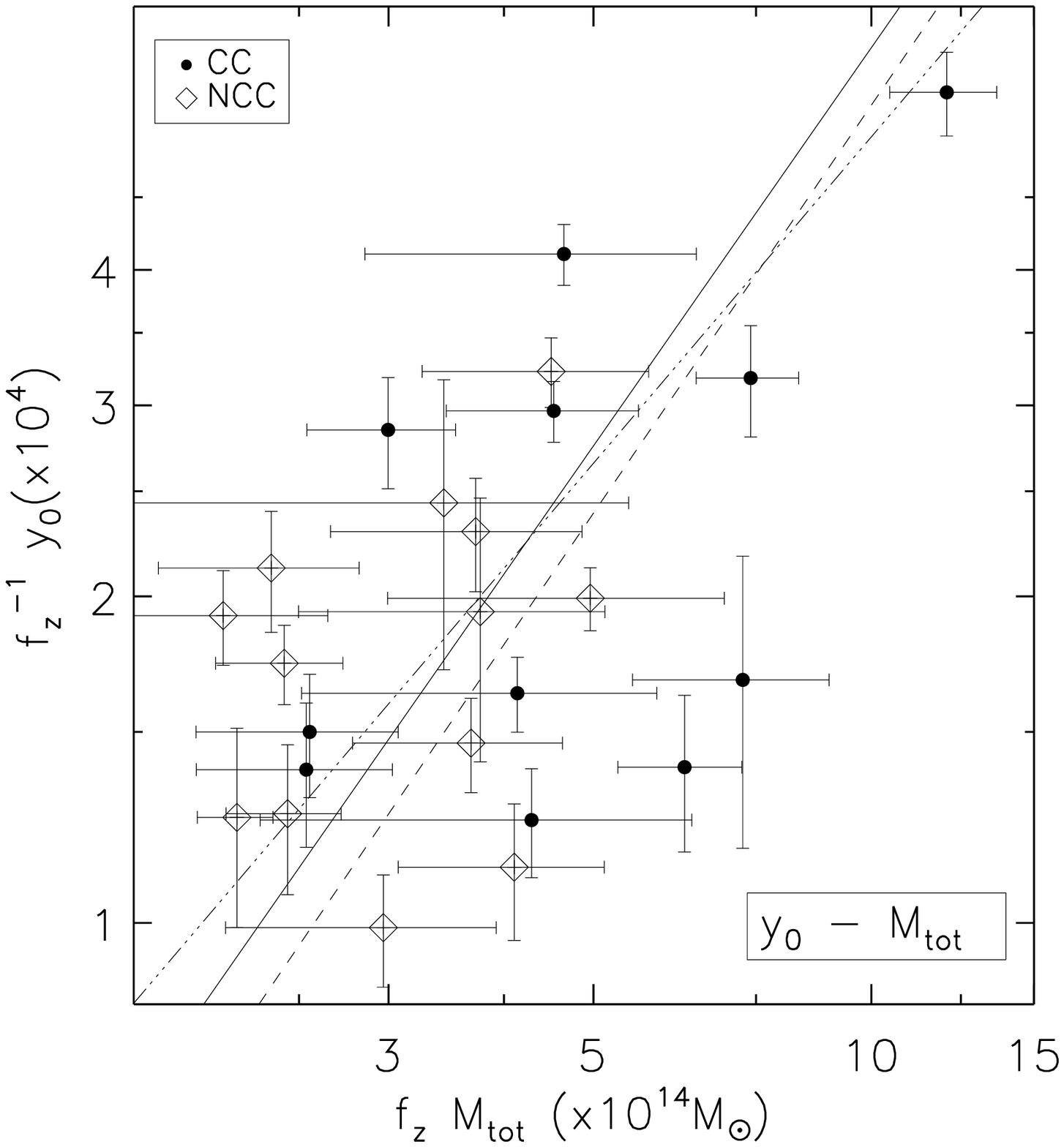,width=0.33\textwidth}
\psfig{figure=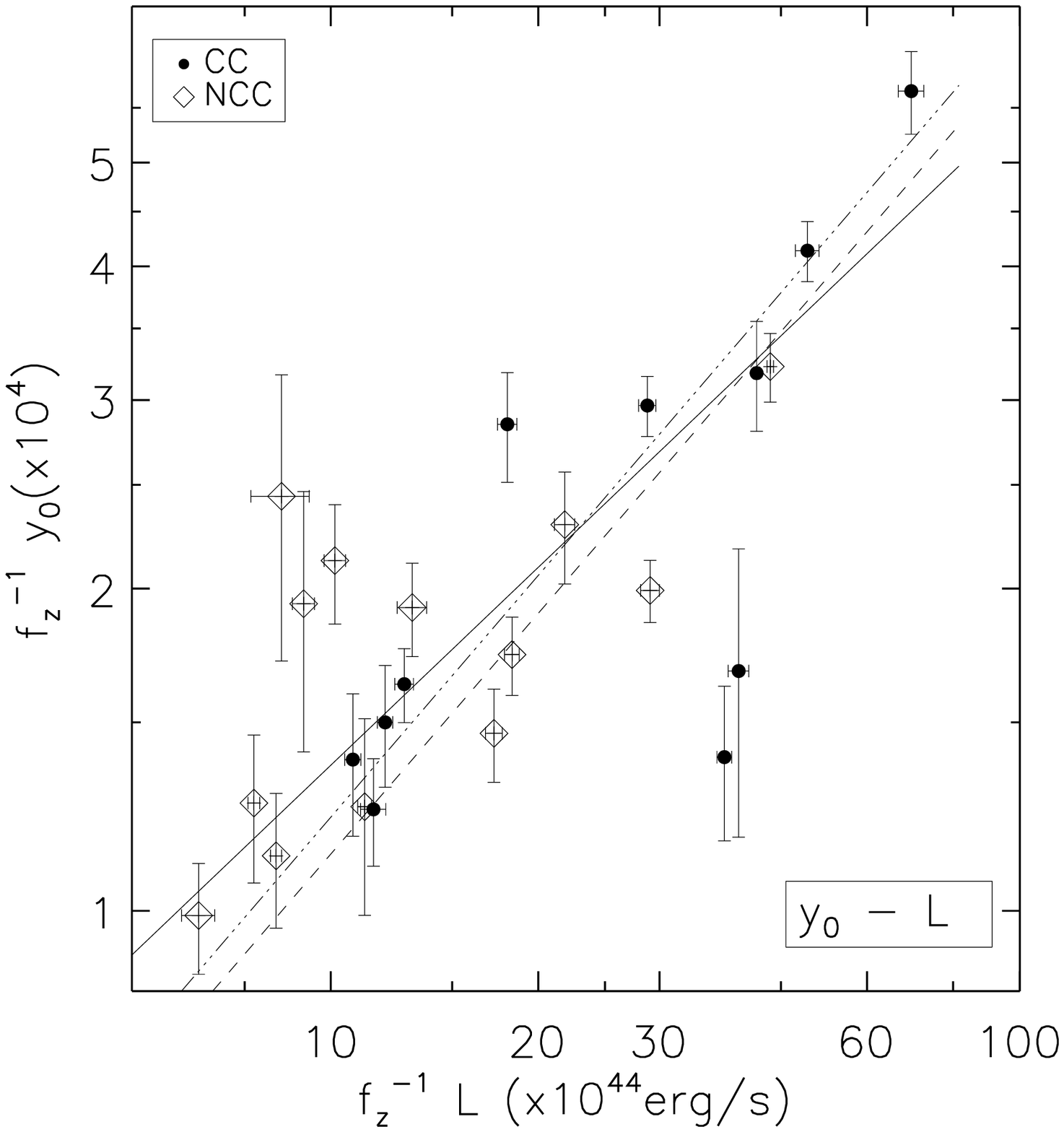,width=0.33\textwidth}
}
\caption[]{As in Fig. \ref{L-T3rela} but for the relations between $y_0-T_{\rm ew}$ (left panel), $y_0-M_{\rm tot}$
(central panel), $y_0-L$ (right panel).}
\label{y_0-rela} 
\end{figure*}

In this section, we discuss first the $Y-y_{0}$ and
$y_{\Omega}-y_0$ relations, which are linking the SZ properties only
(see Fig. \ref{y_0-relaSZ}), and then the relations between SZ and
X-ray quantities (see Fig. \ref{y_0-rela}).  The importance of these
relations relies on the possibility of providing new insights into the
general physical properties of the ICM, in a way complementary to the
X-ray view.  In particular, the different dependence on the gas
density and temperature of the SZ flux ($\sim n_e\,T$) with respect to
the X-ray brightness ($\sim n_e^2\,T^{1/2}$) can allow to reduce some
of the biases present in the X-ray analysis. The presence of
substructures and inhomogeneities in the ICM can indeed strongly
affect some of the X-ray determined physical parameters, like
temperature and luminosity.  An independent approach through the SZ
analysis of some physical quantities can shed more light on the limits
of validity of the ICM self-similar scenario.

\subsubsection{The $Y-y_0$, $y_{\Omega}-y_0$ relations}\label{ydelta-yo1}

For both relations, we find slopes which are smaller than the expected
ones. The discrepancy we measure is larger than the one found by
\cite{2003ApJ...591..526M}.  This is likely due to the fact that the
self-similar model predicts a pressure profile which is steeper than
the observed one: including extra-gravitational energy draws a picture
in which the gas density (and consequently the pressure) has a profile
shallower than the DM density.  This is also confirmed by the
observation that there are differences between CC (which are obviously
more subject to non-gravitational processes) and NCC clusters, having
the former a slightly ($\sim 1 \sigma$) smaller integrated Compton
parameter. We point out that the dispersion in these relations
is very high, probably because of the systematics on the reconstruction of the integrated Compton parameter (see Sect. \ref{thsz0}).

We measure a a strong negative evolution in the $y_{\Omega}-y_0$ relation. 
As pointed as \cite{2003ApJ...591..526M}, this different behaviour of the $y_{\Omega}-y_0$ relation (more in general of the $y_{\Omega}-$X-ray and $y_{\Omega}-$SZ relations) concerning the evolution is likely due to the fact the SZ effect within a fixed angular size samples larger physical region at higher redshifts. This means that the effect of non-gravitational processes are relatively more pronounced if the SZ flux is measured within smaller physical radii, where the density of the ICM is higher: this is expected in a scenario of either preheating, where we can assign a fixed extra-energy per particle, or cooling, where the radiative cooling is more prominent in the denser central regions. This is also in agreement with the general picture emerging by studying entropy profiles \citep{2003MNRAS.343..331P, 2006A&A...446..429P, 2003ApJ...594L..75V, 2001ApJ...546...63T}, which is affected just in the central regions by non-gravitational processes, while the self-similarity is roughly preserved in the halo outskirt, where the dynamics is still dominated by the gravity.

\subsubsection{The $y_0-T$, $Y-T$, $y_{\Omega}-T$ relations}\label{aaw}

We note that $y_0-T$ is the only scaling relation
that deviates by $\ga 3 \sigma$ from the self-similar slope 
(see Table~4 and 5) both when only CC clusters and CC+NCC objects are
considered.  Moreover, we measure an higher normalization in the
CC-only subsample, probably due to the inclusion of the cooling
regions during the SZ data reduction and the subsequent fit in the
visibility plane. These results, in good agreement with the ones
presented in \cite{2004ApJ...617..829B}, are consistently obtained
with both a robust BCES fit and a $\chi^2$-minimization. By applying
the former technique, this relation is also the one that shows the
smaller scatter (both total and intrinsic) around the
best-fit. Furthermore, the $\chi^2$-approach indicates a significant
negative evolution among the 11 CC clusters 
($B^{\rm CC}=-2.12^{+0.96}_{-0.96}$ at $2.5 \sigma$;
$\chi^2_{\rm min} = 23.4$ with 8 d.o.f.) that disappears
when the whole sample of 24 objects is considered. For the NCC sources we do observe hints of positive evolution ($B^{\rm NCC}=0.64^{+0.40}_{-0.40}$): this points to a different behaviour of the cool core and non-cool clusters in the central regions, and different state of relaxation of the gas as suggested by the comparison of the normalization of the fit ($\alpha^{\rm CC}=0.19\pm0.15$ and $\alpha^{\rm NCC}=0.14\pm0.35$).

The best-fitting relations for $Y-T$ 
show a value for the slope in agreement with the value predicted by
the self-similar model either when we consider the CC-only clusters or 
the whole sample, unlike for the $y_0-T$ relation: this is probably due to the
sensitivity of $y_0$ to the cooling region. On the contrary 
for the $y_{\Omega}-T$ 
relation, when we consider the CC clusters, we observe 
a good agreement with the self-similar predictions 
($A^{\rm CC}=1.98\pm0.46$ $A^*=1.50$).

Our results confirms that the $Y-T$ 
relation exhibits a smaller scatter than the $y_0-T$
one, as naively expected. 
Finally we find that the $y_{\Omega}-T$ relation has a larger scatter
than the $Y-T$ one, in contrast with what obtained by
\cite{2003ApJ...591..515M}. Moreover we notice in the CC-only subsample a mildly larger scatter compared to the whole cluster sample.

\subsubsection{The $y_0-M$, $Y-M$, $y_{\Omega}-M$ relations}

These relations show a very good agreement between observed and
self-similar slopes, with a scatter a factor of 2 larger than the
correlation with $T$ (see the previous subsection).  We do not confirm the
low scatter, $S \approx 10-15$ per cent, for the $Y-M$
relation suggested from the numerical simulations by \cite{2006ApJ...650..538N} and
\cite{2005ApJ...623L..63M}: this indicate possible bias in the
determination of $Y$. But it is possible 
that the present simulations are not
completely adequate to reproduce the observed quantities, being the
ICM modeling in hydrodynamical codes quite
complex.

The normalization of the $Y-M_{\rm tot}$ relation has been
investigated in dedicated hydrodynamical simulations to discriminate
between different ICM physics. For example, \cite{2006ApJ...650..538N}
uses non-radiative (NR) and with gas cooling and star formation (CSF)
simulated clusters to find a normalization that varies by about 70 per
cent: for a typical cluster with $M_{2500}=5\times 10^{14} M_{\odot}$, $Y^{NR}=(1.32^{+0.10}_{-0.09})\times 10^{-4}$ and
$Y^{CSF}=(9.01^{+0.78}_{-0.59})\times 10^{-5}$ at $z=0$\footnote{Here we are following his definition of $Y$,
corresponding to $I_0=1$ in eq.(\ref{ydelta-m1}), 
and we adopt his cosmological parameters.}.
At the same mass and overdensity, and fixing the slope to 
the self-similar model, our observed normalization is:
$Y^{\rm CC}=(5.32\pm{1.06})\times 10^{-5}$ and
$Y^{\rm all}=(8.06\pm{1.35})\times 10^{-5}$ for CC-only and all
clusters, respectively. At $\Delta=200$, the observed normalizations
are $Y^{\rm CC}=(1.30\pm{0.74})\times 10^{-5}$ and 
$Y^{\rm all}=(1.22\pm{0.53})\times 10^{-5}$, 
systematically lower than the results in Nagai (2006) ( 
$Y^{NR}=5.13^{+0.57}_{-0.52}\times 10^{-5}$ and 
$Y^{CSF}=3.95^{+0.37}_{-0.34}\times 10^{-5}$) and
more in agreement with the results by \cite{2004MNRAS.348.1401D}, that
measure $Y^{NR}=1.85\times 10^{-5}$, 
$Y^{\rm cool}=1.73\times 10^{-5}$ and 
$Y^{\rm pre-heat}=2.50\times 10^{-6}$ for
non-radiative, cooling (cool) and pre-heating (pre-heat)
simulations, respectively.

We obtain, therefore, that our CC clusters, for which we obtain the most
robust estimates of the total mass at the overdensity of $2500$ (see
Subsect.~\ref{M-Tmwrela1}), well reproduce the distribution measured
in the $Y-M_{\rm tot}$ plane of the objects simulated including
extra physical processes.  Similar conclusions can be drawn for
$Y-T_{\rm mw}$ and $Y-L$ relations.

Finally, we find a negative evolution for the relations under
examinations at $\gesssim 1 \sigma$ confidence level for the 
CC-only clusters (see Table~5).  The
slopes of the correlations tend, however, to deviate from the
self-similar predictions more significantly than the measurements
obtained with the robust fitting technique.  If we fix the slope to
the self-similar value $\mathcal{A^*}$ in these relations between SZ
and X-ray quantities, we still obtain a negative evolution 
at $\approx 1-2 \sigma$ confidence level. We note here that
\cite{2006ApJ...650..538N}, on the contrary, does not find any hint of
evolution in the $Y-M$ relation.

\subsubsection{The $y_0-L$, $Y-L$, $y_{\Omega}-L$ relations}

In general we find a good agreement between the best-fitted slope and
the self-similar prediction.  Compared to other scaling relations, in
these cases the intrinsic scatter is very small ($\sim 0.15$ 
for the $y_0-L$ relation estimated in the CC-only subsample).  We do
not observe significant differences between CC and NCC clusters, being
the estimates of luminosity corrected for the cooling core.

Regarding the evolution, we find suggestions (at $3\sigma$ level) for
a negative evolution in the $y_{\Omega}-L$ relation ($B^{\rm CC}=-1.52^{+0.48}_{-0.44}$). We observe instead positive evolution in the $Y-L$ relation, $B^{\rm CC}=2.40^{+0.44}_{-0.48}$, but negative evolution when we consider the NCC clusters ($B^{\rm NCC}=-0.80^{+0.24}_{-0.20}$).

\section{Conclusions}\label{swqag}

We have presented an analysis of X-ray and SZ scaling relations of a
sample of 24 galaxy clusters in the redshift range 0.14-0.82, selected
by having their SZ measurements available in literature. We have
analyzed the Chandra exposures for these X-ray luminous objects.  We
have reconstructed their gas density,
temperature and pressure profiles in a robust way.  Then, we have
investigated the scaling relations holding between X-ray and SZ
quantities.  By assuming an adiabatic self-similar model, we have
corrected the observed quantities by the factor 
$f_z \equiv E_{z}$, neglecting the factor $\Delta_z$, 
checking that the final results do not change significantly 
in this way: so we can compare our results with the work in the 
literature. We have estimated the values of
normalization, slope, observed and intrinsic scatters, and evolution
to quantify the amplitude of the effects of the non-gravitational
processes in the ICM physics.  In this sense, the combined study of
the SZ and X-ray scaling relations and their evolution in redshift is
a powerful tool to investigate the thermodynamical history in galaxy
clusters.  Indeed, the departures from the self-similar predictions
observed in some of the scaling laws studied in our work confirm
that the simple adiabatic scenario is not wholly adequate to describe
the physics of the X-ray luminous clusters, because it does not
account for a further non-gravitational energy besides the potential
one.  We remind that our results are, by construction, more robust at
$R_{2500}$, where no extrapolation is required and the determinations
of the mass (at least for CC clusters) and the reconstruction of the
integrated Compton parameter are reliable. These results
can be here summarized as follows.

\begin{itemize}
\item 
We observe a good agreement of the normalization of the $M_{\rm
tot}-T$ relation between our results and the ones obtained in
hydrodynamical numerical simulations.  The other X-ray scaling
relations involving a direct propagation of the absolute value of the
measured gas density show a steeper slope than expected from
self-similar predictions. Departures larger than $2 \sigma$ are
observed in the 
$L-T$ ($A^{\rm all}=3.37 \pm 0.39$ vs. $A^*=2$), $L-M_{\rm gas}$
($A^{\rm all}=1.64\pm 0.13$ vs. $A^*=1.33$) and $M_{\rm gas}-T$ ($A^{\rm all}=2.09 \pm
0.23$ vs. $A^*=1.5$) relations.  These results are consistent with
previous analysis on high-$z$ X-ray luminous galaxy clusters
\citep[see, e.g.,][]{2002A&A...391..841E,2005ApJ...633..781K,2006MNRAS.365..509M}.

\item 
Correlations between the investigated SZ quantities and the gas
temperature have the largest deviations from the slope predicted from
the self-similar model and the lowest scatter among similar relations
with different X-ray quantities.  The measured scatter is comparable
to what is observed in the relations between X-ray parameters. The
$Y - T$ relation shows the lowest total and intrinsic scatter both
when CC clusters only and the whole sample are considered.

\item
We observe a strong negative evolution in the $y_{\Omega}-$X-ray and $y_{\Omega}-$SZ relations. A plausible explanation is that the SZ effect within a fixed angular size samples larger physical region at higher redshifts. That means the effect of non-gravitational processes are relatively more pronounced within smaller physical radii.

\item 
The observed normalization of the $Y-M_{\rm tot}$ relation
in cooling-core clusters at $\Delta=2500$, that provide the most
robust estimates of the total masses in our cluster sample, agrees
well with the predicted value from numerical simulations
\citep[see, e.g.,][]{2004MNRAS.348.1401D,2006ApJ...650..538N}.
In particular, we confirm the trend that lower normalization are
expected when some feedback processes take place in the cluster cores:
for a cluster with typical $M_{2500} \approx 5\times 10^{14} M_{\odot}$,
we measure $Y^{\rm CC}=(5.32\pm{1.06})\times 10^{-5}$ in the sample
of CC objects where the cooling activity is expected to be very
effective, and $Y^{\rm all}=(8.06\pm 1.35)\times 10^{-5}$ in the
whole sample.  However, we have to note that the normalization in
hydrodynamical simulations is strictly related to the adopted recipes
to describe physical processes, like gas cooling and star formation.
These processes are also responsible for the production of the cold baryon
fraction, the amount of which is still under debate when compared to
the observational constraints \citep[see,
e.g.,][]{2006MNRAS.367.1641B}.

\item 
The SZ -- X-ray relations are, in general, well described by a
self-similar model parametrized through the dependence upon $f_z$,
when a robust fitting technique, that considers both the intrinsic
scatter and the errors on the two variables, is adopted.  On the
contrary, when an evolution in the form $(1+z)^B$ is investigated by a
$\chi^2$-minimization with error propagations on both $X$ and $Y$
variables, we measure a strong negative evolution at $\ga 1 \sigma$
level of confidence for all relations that involve SZ quantities
($y_0, Y, y_{\Omega}$) and the X-ray measured gas temperature
and total mass.  The slopes of the correlation tend, however, to
deviate from the self-similar predictions more significantly than the
measurements obtained with the robust fitting technique. If we fix
the slope to the self-similar value $\mathcal{A^*}$ in these relations
between SZ and X-ray quantities, 
we obtain stronger hints of negative evolution 
for the $y_0-M_{\rm tot}$ relation ($B^{\rm CC} =-0.88\pm  0.94$) and for the $Y-M_{\rm tot}$ relation ($B^{\rm CC} = -2.30\pm  1.13$).

\end{itemize}

Our results on the X-ray and SZ scaling relations show a tension
between the quantities more related to the global energy of the system
(e.g. gas temperature, gravitating mass) and the indicators of the ICM
structure (e.g.  gas density profile, central Compton parameter
$y_0$).  Indeed, by using a robust fitting technique, the most
significant deviations from the values of the slope predicted from the
self-similar model are measured in the $L-T$, $L-M_{\rm tot}$, $M_{\rm
gas}-T$, $y_0-T$ relations.  When the slope is fixed to the
self-similar value, these relations show consistently a negative
evolution suggesting a scenario in which the ICM at higher redshift
has lower both X-ray luminosity and pressure in the central regions
than the self-similar expectations.  These effects are more evident in
relaxed CC clusters in the redshift range 0.14-0.45, where a more
defined core is present and the assumed hypotheses on the state of the
ICM are more reliable.

A likely explanation is that we need an increase in the central
entropy to spread the distribution of the gas on larger scales: this
could be achieved either by episodes of non-gravitational heating due
to supernovae and AGN \citep[see, e.g., ][]{1991ApJ...383...95E,
1999MNRAS.308..599C, 2001ApJ...546...63T, 2001ApJ...555..597B,
2001ApJ...553..103B, 2002MNRAS.330..329B, 2002MNRAS.336..409B}, or by
selective removal of low-entropy gas through cooling 
\citep[see, e.g.,][]{2001MNRAS.326..649P, 
2001Natur.414..425V, 2002ApJ...572L..19W},
possibly regulated by some mechanism supplying energy feedback
[e.g. the semi-analytical approach proposed by
\cite{2002ApJ...576..601V} and the numerical simulations discussed by
\cite{2002MNRAS.336..527M, 2003MNRAS.342.1025T, 2003MNRAS.343..608K}].

\section*{acknowledgements}
We thank the anonymous referee for a careful reading of the manuscripts 
and suggestions that have improved the presentation of our work. 
We thank Steven Myers for useful discussions on SZ data and NRAO for
the kind hospitality. The visit at NRAO has been partially supported
also by the `Marco Polo' program of University of Bologna.  We
acknowledge the financial support from contract ASI-INAF I/023/05/0
and from the INFN PD51 grant.


\begin{appendix}
\section{Spectral deprojection technique}\label{apecdepte}

The deprojection technique decomposes the observed X-ray emission of
the $i$-th annulus into the contributions from the volume fraction of
the $j$-th spherical shells with $j\le i$, by fixing the spectrum
normalization of the outermost shell to the corresponding observed
values. We can construct an upper triangular matrix
${\mathcal{V}}_i^j$, where the column vectors ${\mathcal{V}}^1$,
${\mathcal{V}}^2$, ...${\mathcal{V}}^n$ represent the ``effective"
volumes, i.e. the volume of the $j$-th shell contained inside the
$i$-th annulus (with $j \ge i$) and corrected by the gradient of
$n_e^2$ inside the $j$-th shell (see Appendix~B for more detail), so
as:
\begin{equation}\label{kk}
K_i \propto \int_{j \ge i} n^2_{\rm e,j}\ dV = 
{\left({\mathcal{V}} \# {\stackrel{\rightarrow}{n_e^2}}\right)}_i\ .
\end{equation}
In the previous equation $\stackrel{\rightarrow}{n_e}\equiv
(n_{e,1},n_{e,2},...,n_{e,n})$, being $n$ the total number of annuli,
having internal (external) radius $r_{\rm in,1}\, ,\, r_{\rm in,2}\, ,
\, ...\, ,  r_{\rm in,n}$ ($r_{\rm out,1}\, ,\, r_{\rm out,2}\, ,\, ...,
r_{\rm out,n} $), with $n\sim 15- 40$; $K_i$ is the MEKAL normalization of the spectrum in the $i$-th annulus; the operator $\#$
indicates the matrix product (rows by columns). Notice that the
integral $\int_{j \ge i} n^2_{\rm e,j} dV$ is of the order of the
emission measure inside the $i$-th ring.\footnote{Hereafter we assume
that the index $j$ ($i$) indicates the shell (ring) of the source of
radius $(r_{\rm in},r_{\rm out})$. } The inversion of this matrix
allows us to determine $n_{e,i}$.

The values of $K_i$ are obtained by rescaling by the observed number
of counts in the $i$-th ring the faked {\it Chandra} spectrum with
absorption, temperature and metallicity measured in that ring.  The
errors are computed by performing 100 Monte Carlo simulations of the
observed counts.  We pointed out that the uncertainties in the
estimates of the projected temperature do not reflect into high
systematic errors in the determination of $K_i$, because of the mild
dependence on $T$ of the cooling function $\Lambda(T)$ integrated in
the considered band ($0.5-5$ keV): $\Lambda(T) \propto T^{-\alpha}$,
with $0.1\la \alpha\la 0.2$ for $T\sim 7- 12$ keV.

This approach is very powerful, because does not require any ``real"
spectral analysis, which could suffer of the poorness of the
statistics and would need at least $\sim$ 2000 net counts per annulus:
we can determine the projected density in annuli even with very small
counts ($\sim 200 - 1000$).  In other words we have an improvement (of
about one order of magnitude) of the spatial resolution in the
spectral analysis.

Concerning the temperature analysis, we have determined its value
$T_j$ in the $j$-th shell, by assuming analytic relations for the mass
density profiles: $\rho=\rho({\bf {r, q}})$, where ${\bf q}=(q_1, q_2,
... \, q_h)$ are suitable parameters.  As discussed in Section~2.4, we
consider two functional forms, a NFW profile with ${\bf q}\equiv
(c,r_{\rm s})$ and a RTM profile with ${\bf q}=(x_{\rm s}^{*},r_{\rm
vir})$.

We performed a spectral deprojection of the observed temperature
$T^*_{\rm shell}$ in a set of $n^*$ annuli with width much larger than
the previous ones, with internal (external) radius $r^*_{\rm in,1}\,
,\, r^*_{\rm in,2}\, ,\, ..., r^*_{\rm in,n^*}$ ($r^*_{\rm out,1}\,
,\, r^*_{\rm out,2}\, ,\, ..., r^*_{\rm out,n^*}$) corresponding to
the ones of the rings in which we have estimated the projected
temperature (see Sect. \ref{sp}), with $n^* \ll n$ ($n^*\sim 3- 8$),
so as to have at least 2000 counts per annulus.  The deprojection
method works in this way:
\begin{equation}\label{aa1}
{\stackrel{\rightarrow}{T^*}_{\rm ring,m}}= {{\left({\mathcal{V^*}} 
\# {\left( {\stackrel{\rightarrow}{T^*}_{\rm shell}} 
\cdot {\stackrel{\rightarrow}{\epsilon^*}} \right)} \right)}_{\rm m}} \; / \, 
{L^*}_{\rm ring,m}\ ,
\end{equation}
where the operator ``$\cdot$" indicates the product: ${\stackrel{\rightarrow}
{T^*}_{\rm shell}} \cdot {\stackrel{\rightarrow}{\epsilon^*}}=({T^*}_{\rm 
shell,1}\,{\epsilon^*_1}, {T^*}_{\rm shell,2}\,{\epsilon^*_2},...{T^*}_{\rm 
shell,n^*}\,{\epsilon^*_{n^*}})$.
In eq.(\ref{aa1}), ${\stackrel{\rightarrow}{\epsilon^*}} =
{\mathcal{V^*}}^{-1} \# {\stackrel{\rightarrow}{L}}_{\rm ring}^*$ is
the emissivity, ${\mathcal{V^*}}={\left[
{\mathcal{V}}^1,{\mathcal{V}}^2, ...{\mathcal{V}}^{n^*}\right]}$,
${L^*}_{\rm ring,m}$ is the luminosity of the $m$-th
ring,\footnote{Hereafter we assume that the index $k$ ($m$) indicates
the shell (ring) having radius $(r_{\rm in}^*,r_{\rm out}^*)$.} and
the generic parameter ${\mathcal{P}}^*$ has the same meaning as above,
but it is evaluated in $n^*$ annuli. The inversion of the matrix in
eq.(\ref{aa1}) allows us to finally estimate the deprojected
temperature $T^*_k$.

We computed the theoretical temperature $T_j$ by numerically
integrating the equation of the hydrostatic equilibrium 
(eq. \ref{aa2}), assuming 
spherical geometry (${\stackrel{\rightarrow}{r}}\equiv r$).
Then we constructed a grid of values for $P_0$ and for the parameters
${\bf q}$ entering the DM density profiles, so as $T_j=T_j({\bf
q},P_0)$. In particular for $P_0$ we have considered the range
\^P$_0-3\sigma_{P_0} \le {P_0} \le $\^P$_0+3\sigma_{P_0}$, where
\^P$_0$ is the expectation value of $P_0$. So we can estimate the
temperature $k T_j({\bf q},P_0)=P({r})/n_{\rm gas}({r})$.

Since the temperature $T_j({\bf q},P_0)$ obtained in this way is given
on a set of $n$ annuli with spatial resolution much better than the
deprojected temperature $T^*_{\rm shell,k}$ defined in the $n^*$
annuli only, we perform a (gas mass-weighted) average to calculate the
temperature $T^{\rm ave}_k({\bf q},P_0)$ in the $k$-th shell:
\begin{equation}
T^{\rm ave}_k({\bf q},P_0) = {{\sum_{r_k \le r_j <r_{k+1}} w_j  
T_j({\bf q},P_0) dV_j } \over {\sum_{r_k \le r_j <r_{k+1}} w_j dV_j }}\ ,
\end{equation}
where $w_j=n_j$ and $dV_j$ represents the volume of the $j$-th shell,
so as to reproduce a mass-weighted temperature.  A
${\chi}^2$-minimization between $T^{\rm ave}_k({\bf q},P_0)$ and
$T^*_{\rm shell,k}$ (with error $\sigma_{_{T^*_{\rm shell,k}}}$),
\begin{equation}
\chi^2= \sum_{k=1}^{n^*} {{ (T^{\rm ave}_k({\bf q},P_0)-T^*_{\rm shell,k})}^2 
/{\sigma^2_{_{T^*_{\rm shell,k}}}}}
\end{equation}
provides us the best estimate of $({\bf q},P_0)$.

We also considered an alternative approach to determine $({\bf
q},P_0)$. Following \cite{2004MNRAS.354...10M}, we perform a weighted
average of $T_k({\bf q},P_0)$ to compute a projected spectral-like
temperature $T_{\rm proj,m}({\bf q},P_0)$ in the $m$-th ring to be
compared with the observed temperature $T^*_{\rm proj,m}$ of the
$m$-th ring:
\begin{equation}
{{T_{\rm proj,m}}}({\bf q},P_0)\!\! =\!\!\! {{{\left(\!\! {\tilde{\mathcal{V^*}}}\! \# \!{\left
( \!{\stackrel{\rightarrow}{T^{\rm ave}}}({\bf q},P_0) \cdot{\stackrel{\rightarrow}{w}\!({\bf q
},P_0)} \right)} \!\! \right)}}_{m} \!\!\!/ {{{\mathcal{L}}}_{\rm ring,m}}},
\end{equation}
where $w_j=n^2_j T_j^{-\alpha}({\bf q},P_0)$, $\alpha=3/4$,
${\stackrel{\rightarrow}{{\mathcal{L}}}_{\rm ring}}({\bf q},P_0)=
{\tilde{\mathcal{V^*}}} \# {\stackrel{\rightarrow}{w}}({\bf q},P_0)$;
${\tilde{\mathcal{V^*}}}={\left[ {\mathcal{V}}^1,{\mathcal{V}}^2,
...{\mathcal{V}}^{n^*},{\mathcal{V}}^{n^*+1},
...,{\mathcal{V}}^{n^*+h}\right]}$ is an extension of the volume
matrix ${\mathcal{V^*}}$ which takes into account the contributions
(up to a distance of 10 Mpc) coming from the $h$ annuli external to
$R_{\rm spat}$.  We have to use the following fitting function, which is a simplified of the functional form of \cite{2005ApJ...628..655V}:
\begin{eqnarray}\label{eq:density:model}
n_e(r) =  \frac{n_0\; (r/r_c)^{-\alpha}
(1+r^\gamma/r_s{}^\gamma)^{-\varepsilon/\gamma}}
{(1+r^2/r_c^2)^{3/2 \, \beta-\alpha/2}}
\end{eqnarray}
with $\gamma=3$, 
and (\ref{aa2}) to extrapolate $n_e(R)$, the pressure and temperature in
regions outside $R_{\rm spat}$.  Notice that the previous definition
of temperature is a very powerful way to remove observational biases:
in fact we are weighting different regions along the line of sight
using different temperatures which are obtained by performing a
spectral fit of a single-temperature model. With this approach we have
a robust determination of the deprojected temperature profile.  The
best estimate of $({\bf q},P_0)$ is obtained through a
${\chi}^2$-minimization between $T_{\rm proj,m}({\bf q},P_0)$ and
$T^*_{\rm proj,m}$ :
\begin{equation}\label{chi2ww}
\chi^2= \sum_{m=1}^{n^*} {{ (T_{\rm proj,m}({\bf q},P_0)-T^*_{\rm proj,m})}^2 
\over{\sigma^2_{T^*_{\rm proj,m}} + \sigma^2_{T_{\rm proj,m}} }}\ .
\end{equation}
Here $\sigma^2_{T_{\rm proj,m}}$ accounts for the statistical errors
in eq.(\ref{aa2}) coming from the measured errors for $n_{\rm
gas}(r)$.  The reduced $\chi^2$ resulting from this method is better
than in previous case: this is likely due to the fact that the
deprojected temperature $T^*_{\rm shell}$ strongly relies on
assumptions, like spherical symmetry and uniform density profile,
which are not completely satisfied in real clusters.  Moreover the
values of $T^*_{\rm shell,k}$ are not independent: in fact we relate
the deconvolved temperature, gas density and spectra normalization of
the outermost shell to its observed values and then we compute the
physical parameters in the $m$-th annulus by opportunely accounting
for the contributions of the $k$-th shell ($k\ge m$): this could
propagate possible systematic errors from the external regions, where
the determination of the physical properties cannot be so adequate
because of the bad statistic. All the deprojected quantities presented
in the present work refer to the second approach ($T_{\rm proj}$)
only.

\section{Determining the Effective volume}

  \renewcommand{\thefigure}{A-\arabic{figure}}
  \setcounter{figure}{0}  
 
\cite{1983ApJ...272..439K} computed the geometrical volume of the $j$-th shell 
intercepted by the $i$-th annulus (with $j \ge i$) as:
\begin{equation}
{{V}}_i^j \!\!=4 \pi \!\int_{r_{\rm in_i}}^{r_{\rm out_i}} d\,r \;r
\!
\int_{(r_{\rm out_j}^2-r^2)^{1/2}}^{(r_{\rm in_j}^2-r^2)^{1/2}} d\, z\ .
\end{equation}

Notice that when we use a geometrical volume to deproject the physical
parameters (like the density as example) we are assuming that they are
nearly constant in the shell.  This introduces a systematic bias in
the deprojected quantity, that tends to be increased when the gradient
of the physical parameter is not negligible or when the rings are
wide. McLaughlin (1999) partially corrected this bias by referring the density to an average radius, $r_{\rm ave}\equiv{((r_{\rm out}^{3/2}+r_{\rm 
in}^{3/2})/2)}^{2/3}$.

Here we introduce a new definition of the volume, the effective volume
${\mathcal{V}}$, which takes into account the real gradient of the
physical parameters as a function of the radius. We assumed that we
are weighting the unknown physical parameter $\mathcal{P}$ in the
$j$-th shell using a function $w(R)$, whose gradient is only due to
the case of the squared density ($w(R)\propto n_e^2$).

We have modeled the density inside the $j$-th shell as a local
power-law, $n(R)=n_{e,j}\,
{f(R)}^{-\alpha}$, where $f(R)=(R/r_{\rm ref_j})$,
$r_{\rm ref_i}\equiv (r_{\rm in_j}+r_{\rm out_j})/2$,
$\alpha(R)=-\log{ (n^{j+1}/n^j) } / \log{(r_{\rm ref_{j+1}}/r_{\rm
ref_j})} + {\mathcal{O}}(\alpha)$. 
We first calculated $\alpha(R)$ by relying on the
initial density obtained from the geometric volume-deprojection on a
radius $r_{\rm ref_i}$: in this way the introduced errors on $\alpha$
are negligible (${\mathcal{O}}(\alpha)$).

We define $r$ as the projection of $R$ on the sky plane, with
$R^2=r^2+z^2$, being $z$ the distance along the line of the sight. So,
if $n(r_{\rm ref_j})$ is the density in the $j$-th shell, the observed
parameter $\mathcal{P^*}$ is related to the theoretical one by:
\begin{eqnarray}\label{P}
{\stackrel{\rightarrow}{P^*}}= \int dV {\stackrel{\rightarrow}{P}}\;
w(R)= \int dV {\stackrel{\rightarrow}{P}}\; n^2_{\rm e}(R) = &
\nonumber\\ ={ {\left({\int dV {f(R)}^{-2\alpha}} \right)} \# \left({
{\stackrel{\rightarrow}{n_{e,j}}}^2 {\stackrel{\rightarrow}{P}} }
\right)}= {\mathcal{V}} \# {\left({
{\stackrel{\rightarrow}{n_{e,j}}}^2 {\stackrel{\rightarrow}{P}}
}\right)}\ .
\end{eqnarray}
So we can re-write the effective volume ${\mathcal{V}}$ as:
\begin{eqnarray}
{\mathcal{V}}_i^j\!=\!\!\!{\int_{j\ge i} \!\!\!\!\! dV
{f(R)}^{-2\alpha}}\!\!= \!\!4 \pi \! \int_{r_{\rm in_i}}^{r_{\rm
out_i}}\!\!\!\!\!\!\!\!\!\!\!  d\,r \;r \!  \int_{(r_{\rm
out_j}^2-r^2)^{1/2}}^{(r_{\rm in_j}^2-r^2)^{1/2}} d\, z \; f(R)^{-2\alpha}\ .
\end{eqnarray}

The effective volume ${\mathcal{V}}_i^j$ is equal to the geometric one
${{V}}_i^j$ if $\alpha =0$, i.e. when we have negligible gradients of
$n(R)^2$ in the $j$-th shell. This is approximately true only in the
case in which we have a good spatial resolution, for example when we
consider $n$ annuli ($n \sim 15- 40$) in the brightness image (see
Section \ref{depr}). But this is false when we have $n^*$ annuli, with
$n^* \ll n$ ($n^*\sim 3- 8$) in the spectral analysis, for which a
larger statistics (at least $\sim$ 2000 net counts per annulus) is
required.  In this last case, for example, it is possible to
underestimate the true density in the external regions by $5-10$ per
cent by using the geometrical volume instead of the effective one:
this corresponds to set ${\stackrel{\rightarrow}{P}}$ equal to
${\mathcal{I}}$ (being ${\mathcal{I}}$ the identity matrix) in
eq.(\ref{P}), and ${\stackrel{\rightarrow}{P^*}}
\propto K$ (see eq. \ref{kk}).  The analysis we performed shows that 
the case in which adopting the effective volume is important is in
eq.(\ref{aa1}): using an effective volume can avoid to introduce
systematic errors in the determination of the cluster masses.

\end{appendix}

\label{lastpage}
\end{document}